\documentclass[fleqn,usenatbib]{mnras}
\usepackage{enumitem}
\usepackage{changes}

\usepackage[T1]{fontenc}

\DeclareRobustCommand{\VAN}[3]{#2}
\let\VANthebibliography\thebibliography
\def\thebibliography{\DeclareRobustCommand{\VAN}[3]{##3}\VANthebibliography}

\usepackage{graphicx}	% Including figure files
\usepackage{amsmath}	% Advanced maths commands
\usepackage{amssymb}	% Extra maths symbols
\title[\OurModel]{\OurModel: Machine-learning based pipeline to identify and evaluate planet candidates from TESS}

\author[S. Rao et al.]{
Sriram Rao$^{1}$,
Ashish Mahabal$^{2,3}$,
Niyanth Rao$^{4}$,
and Cauligi Raghavendra$^{5}$
\\
$^{1}$LinkedIn Corporation, Mountain View, CA 94043, USA\\
$^{2}$Division of Physics, Mathematics, and Astronomy, California Institute of Technology, Pasadena, CA 91125, USA\\
$^{3}$Center for Data Driven Discovery, California Institute of Technology, Pasadena, CA 91125, USA\\
$^{4}$Saratoga High School, Saratoga, CA 95070, USA \\
$^{5}$University of Southern California, Los Angeles, CA 90089, USA
}

\date{Accepted 2021 January 13. Received 2020 December 28; in original form 2020 October 9.}

\pubyear{2020}

\newcommand{\DAVE}{\textsc{dave}\ }

\newcommand{\AUCScore}{0.908}

\newcommand{\InjectionRecall}{91.9\%}
\newcommand{\InjectionKP}{91 of 99}

\newcommand{\TestPrecision}{88.8\%}
\newcommand{\TestRecall}{74.3\%}
\newcommand{\TestAccuracy}{87.2\%}

\newcommand{\PCprob}{0.7}
\newcommand{\PCprobNew}{0.4}
\newcommand{\Wotan}{\textsc{W\={o}tan}}

\newcommand{\CombinedTestPositive}{19.1\%}  % (98 + 56) / 807.  56 is from the test set we originally had.
\newcommand{\NewCands}{38}
\newcommand{\OurModel}{Nigraha}
\newcommand{\TIC}[1]{\texttt{TIC {#1}}}

\makeatletter
\definecolor{ForestGreen}{RGB}{0,0,0}
\@namedef{Changes@AuthorColor}{ForestGreen}
\colorlet{Changes@Color}{ForestGreen}
\makeatother

\usepackage{newtxtext,newtxmath}

\begin{document}
\label{firstpage}
\pagerange{\pageref{firstpage}--\pageref{lastpage}}
\maketitle

\begin{abstract}
The Transiting Exoplanet Survey Satellite (TESS) has now been operational for a little over two years, covering the Northern and the Southern hemispheres once. The TESS team processes the downlinked data using the Science Processing Operations Center pipeline and Quick Look  pipeline to generate alerts for follow-up. Combined with other efforts from the community,  over two thousand planet candidates have been found of which tens have been confirmed as planets. We present our pipeline, \emph{\OurModel}, that is complementary to these approaches.  \OurModel\ uses a combination of transit finding, supervised machine learning, and detailed vetting to identify with high confidence a few planet candidates that were missed by prior searches. In particular, we identify high signal to noise ratio (SNR) shallow transits that may represent more Earth-like planets. In the spirit of open data exploration we provide details of our pipeline, release our supervised machine learning model and code as open source, and make public the \NewCands\ candidates we have found in seven sectors. The model can easily be run on other sectors as is. As part of future work we outline ways to increase the yield by strengthening some of the steps where we have been conservative and discarded objects for lack of a datum or two.
\end{abstract}


%%%%%%%%%%%%%%%%%%%%%%%%%%%%%%%%%%%%%%%%%%%%%%%%%%

%%%%%%%%%%%%%%%%% BODY OF PAPER %%%%%%%%%%%%%%%%%%

\section{Introduction}
\label{sec:intro}
Launched in 2018, the \textit{TESS} (Transiting Exoplanet Survey Satellite) mission \citep{Ricker_TESS} is an all-sky digital survey aimed at
observing planetary transits of their host stars.   
The TESS spacecraft monitors $24^{\circ} \times 96^{\circ}$ sectors of the sky for contiguous periods of 27 days apiece; data is downlinked once every 13.5 days for further processing.   TESS monitors the sky in an anti-Sun direction, allowing prompt ground-based followup if the candidates are identified quickly. For context, the data for each sector comprises approximately
20,000 light curves.  On average, less than 1\% of these are potential planetary candidates for follow-up.  Effectively, the number of
planet candidates are dwarfed by a vast majority of false positives (i.e., variable stars, eclipsing binaries, etc.) and/or are affected by instrument systematics
which cannot be filtered out.

Classical machine learning based methods such as decision trees, $k$-nearest neighbors, and random forests have been explored for classifying light curves \citep{Thompson_2015, McCauliff_2015, Armstrong_2016, 2018MNRAS.478.4225A}.   
Of these systems, the \emph{Autovetter} \citep{McCauliff_2015}, which is a random forest classifier that makes decisions based on the metrics generated by the Kepler pipeline,
has been successfully used  to make initial dispositions for Kepler/K2 candidates.
Convolutional neural networks (CNN) which are effective at image recognition tasks have been explored as an alternate approach 
for classifying light curves \citep{Pearson_2017, Zucker_2018, Shallue_2018, Ansdell_2018, Dattilo_2019, 2019MNRAS.483.5534S, Yu_2019, Osborn_2020}.  These models 
use supervised learning with a CNN where the neural network learns the characteristics
from light curves.
At a high level these systems
employ the following procedure for rapid identification of targets:
%\begin{itemize}
(1) Pre-process the light curves to search for transits (i.e., periodic dips in the flux) to derive transit parameters such as period and depth,
(2) from the transit parameters build phase-folded light curves which are subsequently used to train a CNN-based model with labeled data, and
(3) make predictions on newly observed data.
%\end{itemize}
This procedure has been successfully used in practice with Kepler and K2: (1) \cite{Shallue_2018} use their \texttt{Astronet} deep learning model on Kepler data to identify several candidates from
which they validated two new exoplanets and (2)  \cite{Dattilo_2019} have adapted \texttt{Astronet} model for K2 data (dubbed, \texttt{Astronet-K2}) to discover two new exoplanets.
\cite{Ansdell_2018} extend the basic \texttt{Astronet} model to include stellar parameters and demonstrate with Kepler data that their \texttt{Exonet} model improves model accuracy by using
domain information.  

In this paper, we explore the use of CNNs for classifying TESS light curves to search for planet candidates.  A natural starting point is these existing models based on Kepler data.
%{CNNs are proving to be particularly fruitful for finding planet candidates. Therefore, these models represent a natural starting point for classifying TESS data.} 
However, the 
systematics with individual TESS sectors are significantly different when compared to Kepler/K2 (viz., 27 days of data versus 4 years/81 days of data).  Hence, prior deep-learning approaches for classifying TESS data such as,
\cite{Yu_2019},  \cite{Osborn_2020} adapt and modify \texttt{Astronet} and \texttt{Exonet}, respectively, and identify new candidates. %\deleted{from TESS data}. 
Our work is similar in spirit in that we start with
a hybrid of these two models and develop novel extensions that allow us to explore different aspects of the input parameter space to identify candidates.  

The core contribution of this paper is in the design and evaluation of a deep learning based pipeline for identifying planetary candidates from TESS data.
In building our pipeline, \emph{\OurModel}\footnote{Sanskrit word to mean \emph{determined} with part of the word meaning a planet.}, two aspects are novel, namely the input representation of the light curve to the CNN model (\S~\ref{sec:data}) and the model's architecture.  
We train and evaluate the model using labeled data from actual TESS sectors (\S~\ref{sec:models}).  
Our model has an AUC score of \AUCScore. 
We also evaluate our model by augmenting our test set with bona-fide planets and find that it recovers \InjectionRecall\ (\InjectionKP)  (Figure~\ref{fig:kp-cdf}).
We apply our trained model to make predictions on data from TESS Sector 6 and Sectors 21---26  and find new candidates (\S~\ref{sec:candidates}).  We then
perform preliminary vetting on a subset of these candidates using a series of steps including \texttt{DAVE} \citep{Kostov_2019_DAVE}, Gaia RUWE, and 2MASS to identify a promising sample of \NewCands\ new candidates for follow-up (Table~\ref{tab:meta-vetted-pcs-1}, 
Table~\ref{tab:meta-vetted-pcs-2}, and Table~\ref{tab:meta-unvetted-pcs}).  
Finally, \S~\ref{sec:conclus} offers concluding remarks and presents directions for future work.

% next section : data
\section{Data Preparation}
\label{sec:data}
In this section, we describe the data preparation steps for generating input representation of light curves to the machine learning models.
We describe our procedure for searching for transits in light curves (\S~\ref{subsec:transit-search}), and then use that for
building the input representation (\S~\ref{subsec:input-rep}).  We begin by describing the labeled datasets used in this paper.  

\subsection{Datasets}
We use pre-processed TESS light curves downloaded from Mikulski Archive for Space Telescopes (MAST) \footnote{\url{http://archive.stsci.edu/tess/bulk_downloads/bulk_downloads_ffi-tp-lc-dv.html}}.
These have a two-minute cadence over a 27-day period and have instrumental systematics removed.
  We use the Pre-Search Data Conditioning Simple Photometry flux (\texttt{PDCSAP\_FLUX}) time series from each light curve to search for \emph{threshold crossing events} (TCEs). These are periodic flux variations  potentially caused by exoplanets transiting their host star.  
We train our models using TCEs 
from various TESS sectors and then use the trained models to identify new candidates.

\subsection{Training Labels}
We use two sets of labeled data for this work.  We use the TESS TOI catalog\footnote{\url{https://exofop.ipac.caltech.edu/tess/}} to construct our first labeled data set (\textbf{DS1} in Table~\ref{tab:TESS-TCEs}).  This catalog
specifies the TESS identifier for which an alert was generated (and the associated light curve is available from MAST), the disposition for that target, and additional metadata.
With TESS labels, \texttt{KP} corresponds to planets known prior to launch, \texttt{CP} are confirmed planets originally identified using TESS data, and \texttt{PC} are
targets that have been identified as planet candidates.
Note that the TESS TOI catalog contains alerts/labels for targets in all TESS sectors to date.  The catalog is therefore a ``living'' document.  When the first draft of this paper was complete, the TOI catalog had labels from Sectors 1---26.  From this data set we set aside
labeled targets from sectors on which we make predictions.  That is, in this paper we train our models and then generate predictions for planetary candidates
in Sector 6/Sectors 21---26.   The second labeled data set is from \cite{Yu_2019} (\textbf{DS2} in Table~\ref{tab:TESS-TCEs}) which
they have made publicly available\footnote{\url{https://github.com/yuliang419/Astronet-Triage/blob/master/astronet/tces.csv}} and we use without any modifications.
Their labels are for targets observed from TESS sectors 1--5. 
For TCEs common in the two datasets, we use the labels from \textbf{DS1} as ``ground truth''.  
The dispositions for the various TCEs and the datasets from which they were obtained is shown in Table~\ref{tab:TESS-TCEs}.  Note that the dataset is imbalanced: 9.8\% correspond to positive labels.

\begin{table}
    \centering
    \begin{tabular}{l|r|r|c|c}
    \hline
    
      \textbf{Type}  & \multicolumn{2}{c|} {Count} & \multicolumn{2}{c|} {Label}\\
    & \textbf{DS1} & \textbf{DS2}  &\textbf{P} & \textbf{N} \\   \hline
        Known Planet (KP) & 166 & & \checkmark \\
    Confirmed Planet (CP) & 38 & & \checkmark \\
    Planet Candidate (PC) & 274 &  165 & \checkmark \\
    Eclipsing Binary (EB) & & 711 & & \checkmark \\
    False Positive (FP) & 139 & & & \checkmark \\
    Instrumental Noise (IS) & & 520 & & \checkmark \\
    Variable Star (V) & & 656 & & \checkmark \\
    Other (O) & 1 & 1 & & \checkmark \\
    Junk (J) & & 3917 & & \checkmark \\ \hline
   % Total & 618 & 5970 \\ \hline
   Total & 618 & 5970 & 643 & 5945 \\ \hline
    Grand Total & \multicolumn{4}{c} {6588} \\ \hline
    \end{tabular}
    
    \caption{Training labels for the TCEs from the two data sets used in this paper. In the labels column, P indicates positive labels, and N, the negative labels.  
    }
    \label{tab:TESS-TCEs}
\end{table}

\subsection{Transit Search} 
\label{subsec:transit-search}

\begin{figure*}
 \begin{center}
\includegraphics[width=\textwidth]{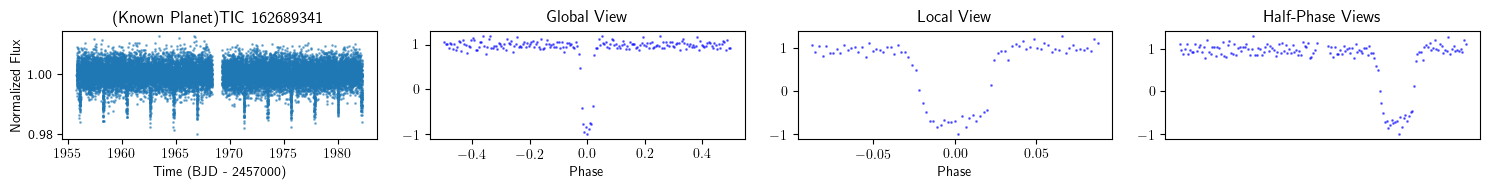}
\includegraphics[width=\textwidth]{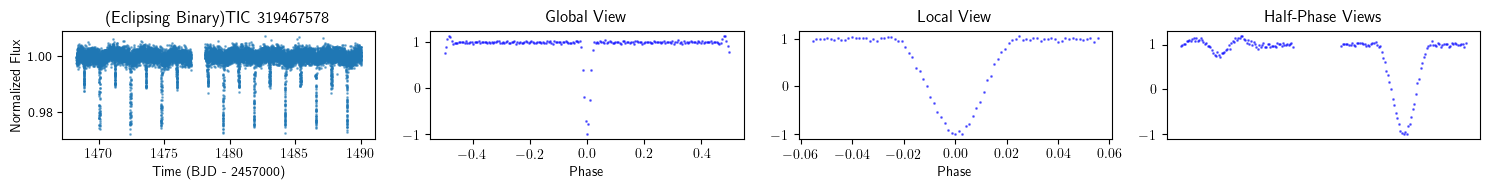}
\includegraphics[width=\textwidth]{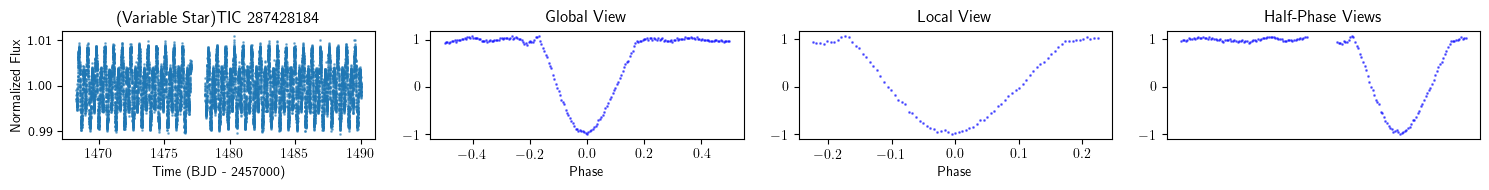}
\includegraphics[width=\textwidth]{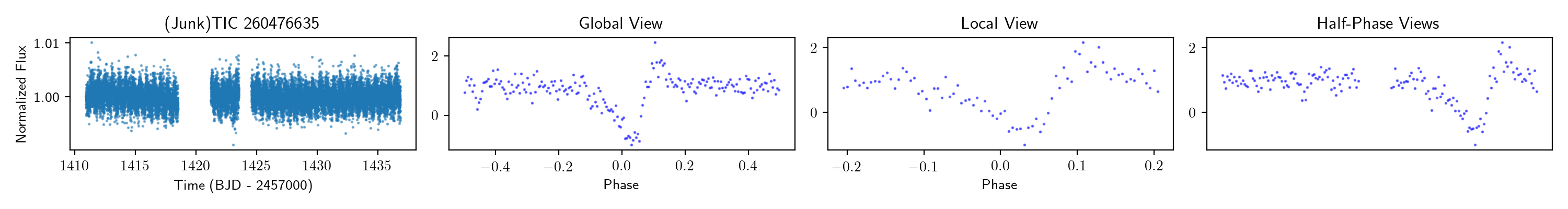}
\includegraphics[width=\textwidth]{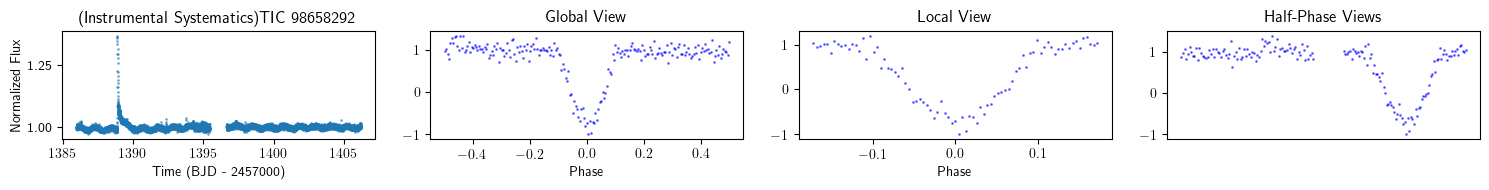}		
\caption{Sample views for the various classes of TCEs that are input to our binary model.
The classes, in order from top to bottom, are \texttt{KP}, \texttt{EB}, \texttt{V}, \texttt{J}, and \texttt{IS}.  
The y-axes for the three views are normalized fluxes. Only KPs are used as positive labels in our model, and the rest as negative labels.
}
\label{fig:all-views}
\end{center} 
 \end{figure*} 

Searching for transits in a stellar light curve entails computing a periodogram from which transit parameters such as, period $(P)$, duration $(t_{dur})$, 
and epoch ($T_0$ or mid-point of transit) are obtained.
Traditionally, Box Least Squares \citep[BLS;][]{Kov_cs_2002} has been the widely used algorithm for this purpose. It builds a periodogram
by searching for  box-like transits in a given light curve.   Recently, Transit Least-Squares \citep[TLS;][]{Hippke_2019} has been proposed as 
an alternative to BLS with the objective of improving the detection of shallow transits, i.e., sensitivity towards detection of transits caused by near-Earth sized planets.  
The key idea in TLS is to build an analytical model of the transit with stellar limb darkening \citep{Mandel_2002}.  \cite{Hippke_2019} point out that
TLS is able to detect shallow transits at an acceptable computational cost (i.e., approximately 10 seconds per TESS lightcurve when run on
an 8-core Intel i9 laptop).  Furthermore, the false positives  produced by TLS are comparable in number with BLS when detecting transits.
When applied to Kepler/K2 data, %subsequent papers
\cite{Heller_2019_1,Heller_2019_2} show that TLS is able to discover Earth-size planets that were previously missed.  That is, BLS did not recover the transits of a sub-Earth in a multi-planet system on {\tt K2SFF} detrended lightcurves; in contrast, TLS did.
These results from TLS are particularly interesting in the context of TESS: the candidate set of planets targeted by TESS include 
super-Earths orbiting close to their host star for which the transits are necessarily shallow.  Motivated by these observations, we
use TLS for transit detection.  

For all the datasets used in this paper, we compute transit parameters using the publicly available open source version of TLS ({\tt TLS version 1.0.25}).
As a preprocessing step we detrend the \texttt{PDCSAP\_FLUX} time series from each TESS light curve file
using an approach similar to \cite{Heller_2020}.  
We remove variability on timescales longer than the maximal expected transit duration with Tukey's biweight filter using
the \Wotan\ software package \citep{Hippke_2019_Wotan}.  \Wotan\ provides a convenience API for estimating the transit duration
as a function of mass, radius, and period.  To this API, we input mass, radius estimates from the TESS input catalog \citep{Stassun_2018} and as a prior, we set the
maximum period to 13.5 days (i.e., half the duration of the observing period of a TESS sector).  We then use a
running window of three times this duration for detrending the light curve.
For a more detailed discussion on the choices of the filter and the associated parameters, we refer the reader to Section 2.1 of \citet{Heller_2020}.
Subsequently, we use TLS APIs to compute the power spectrum on the detrended data.  For this step, we input stellar limb darkening parameters for TESS targets
as estimated by \cite{Claret_2018} and use default values for the remaining parameters in TLS APIs (see documentation\footnote{\url{https://transitleastsquares.readthedocs.io/_/downloads/en/latest/pdf/}}).
Based on these inputs, TLS  computes the transit parameters corresponding to the best fit for the  data.  
Additionally, 
TLS also returns the number of transits, signal-to-noise ratio (SNR) as defined by \cite{Pont_2006}, and 
the signal detection efficiency (SDE) which represents the peak in the TLS power spectrum.
Using these values, for a light curve to contain a TCE, we require (a) at least 2 transits, (b) an SNR $> 7.1$, and (c)
 an $SDE_\text{TLS} > 9$.  
 
\begin{figure}
\begin{center}
\includegraphics[width=\columnwidth]{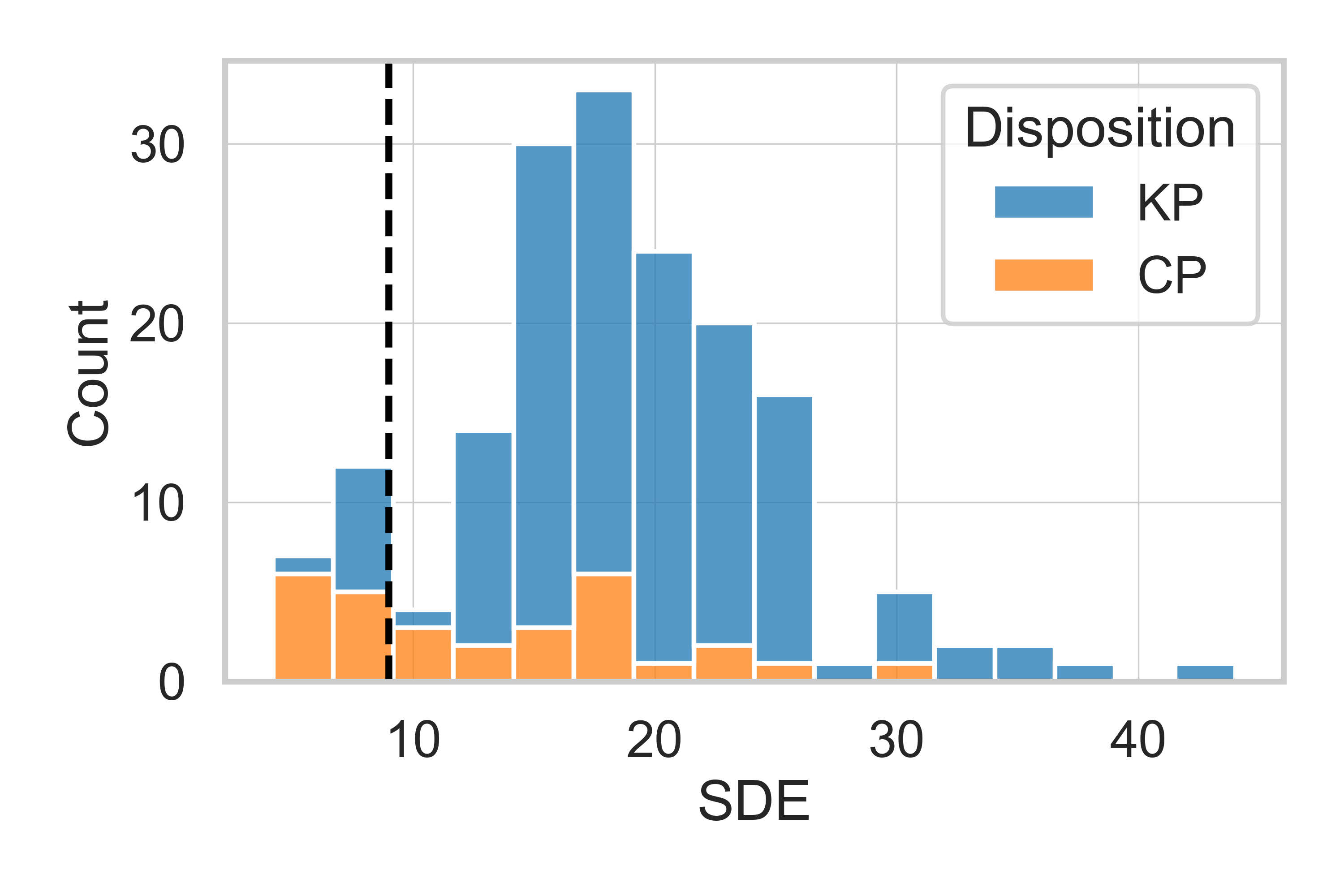}
\caption{Distribution of the TLS computed SDEs for the various \texttt{KP, CP} from the TOI catalog.  Our choice of a threshold for a TCE ($SDE_{\text{TLS}} > 9$, corresponding to the vertical dotted line) is conservative and hence, we fail to recover 29 of these TCEs.} 
\label{fig:tce-sde-dist}
\end{center} 
\end{figure}

 The threshold applied on $SDE_\text{TLS}$ impacts TLS's false positive rate of recovering transits.  It represents a trade-off between the false positive rate and the number of TCEs that are recovered by the transit search.  
 Our choice of  $SDE_\text{TLS} > 9$ is based on the TLS false positive rate of recovering transits to about $10^{-4}$ in the limiting case of white noise (see Section 3.1 of \citet{Heller_2019_1}).  Of the 1168 light curves we searched from the TOI catalog, we recover 618 TCEs that are in \textbf{DS1}. An interesting observation is that our procedure recovers vast majority of the TCEs with a final disposition of \texttt{KP, CP} (see Figure~\ref{fig:tce-sde-dist}).  Our choice for $SDE_\text{TLS}$ threshold is intentionally conservative since machine vetting for TESS candidates is presently limited (see \S~\ref{subsec:vetting}). With even better vetting,  we could, in future, lower the SDE threshold to recover more of the \texttt{KP}s, \texttt{CP}s  (and \texttt{PC}s).
   
 %Our choice of $SDE_\text{TLS}$ biases this trade-off towards lowering the false positive rate.  That is, at $SDE_\text{TLS} > 9$ the TLS false positive rate of recovering transits is about $10^{-4}$ in the limiting case of white noise (Section 3.1 of \citet{Heller_2019_1}). We illustrate these effects by evaluating our transit search procedure for constructing \textbf{DS1}.  Of the 1168 light curves we searched, we recover 618 TCEs from the TOI catalog.  We find that our procedure recovers vast majority of the TCEs with a final disposition of \texttt{KP, CP, FP}. Figure~\ref{fig:tce-sde-dist} shows the distribution of $SDE_\text{TLS}$ for those TCEs. Note that nearly 30 TCEs with an $SDE_\text{TLS} < 9$ to the left of the dotted vertical line are not recovered.

Besides computing $P, t_{dur}, T_0$, TLS also returns several other
transit parameters that we use and they are described below (see \S~\ref{subsec:stellar-transit}).

%\deleted{Using these values, for a light curve to contain a TCE, we require at least 2 transits, an SNR $> 8$, and an SDE $ > 7$.  
%Note that at a threshold of SDE $ = 7$ the TLS false positive rate of recovering transits is about 1\% in the limiting case of white noise and is same as that of BLS (Section 3.1 of \cite{Heller_2019_1}).
%Since our pipeline evaluates TCEs, as part of this pre-processing step, we discard any light curves that do not contain TCEs.}

%%%%%%%%%%%%%%%%%%%%%%%%%%%%%%%%%%%%%%%%%%%%%%
\subsection{Input Representation}
\label{subsec:input-rep}

We use the \textsc{lightkurve} toolkit \citep{2018ascl.soft12013L} to prepare the light curves as input to our machine learning model (see \S~\ref{sec:models}).  Our method is similar
to \cite{Shallue_2018}.  We
iterate over each light curve containing a TCE and perform the following sequence of steps.   First, we clean the light curve to remove outliers
that are at least $5 \sigma$ above the median.  
Second, we remove low frequency trends from the light curve on timescales longer than three times the expected transit duration
using the Savitzky-Golay filter.  We use \textsc{lightkurve}'s \texttt{flatten()} API.  
To prevent the in-transit points from being detrended out, from the mid-point of each transit in the complete light curve, we mask out points that are within $-1.5 \cdot t_{dur}$ to $1.5 \cdot t_{dur}$.
We use default values for the remaining parameters\footnote{\texttt{window\_length = 101, polyorder = 2,  break\_tolerance = 5, niters = 3, sigma = 3}} of this API (see documentation\footnote{\url{https://docs.lightkurve.org/}}).
Third, we phase-fold the detrended light curve at the epoch (i.e., $T_0$ that we computed above) so that the
mid-point of the transit in all the light curves is aligned.  
Finally, we bin the phase-folded light curves to construct two views: (1) a ``global view'' with 201 bins which
corresponds to a transit event for a single period (i.e., the phase between -0.5 and 0.5), and (2) a ``local view'' with 81 bins which 
is a "zoom" in view of the transit between $-2.0 \cdot t_{dur}$ to $2.0 \cdot t_{dur}$.  
In Figure~\ref{fig:all-views}, the columns labeled ``Global view'' and ``Local view'' show sample representation of the respective Global/Local views for the various classes of lightcurves. 

With just the Global/Local views as input, a neural network can be trained to classify whether a light curve contains a transit.  However, further separation of eclipsing binaries
from genuine planet candidates is non-trivial.  This is partly because the neural network's input data representation 
is focused on the characteristics of a single transit event and the shape of that transit event in the light curves of those two systems is quite similar.  
Hence, to improve model performance, we incorporate, with certain modifications, additional secondary views as input as suggested in \citet{Shallue_2018}.

Our approach for constructing a secondary view is based on the characteristics of light curves of eclipsing binaries: The light curve contains a secondary
eclipse whose location is dependent on the system's orbit.  Specifically, for eclipsing binaries in a circular orbit, the secondary eclipse occurs at half the period (i.e., midway between
a pair of consecutive primary transits).  Based on this observation, we build an auxiliary input representation that enables our model to prune a class of eclipsing binaries from
planet candidates.   We construct a secondary view (called ``Half-Phase'' view) 
that is similar to the local view (described above), but
with a few changes: (1) we bin and phase-fold the lightcurve at $T_0$ (i.e., phase = 0), and
(2) we bin and phase-fold the lightcurve at 
$T_0 + 0.5 \cdot P$ (i.e., phase = 0.5),
and (3) we concatenate these two and normalize them as before.  
We illustrate these ideas in Figure~\ref{fig:all-views}.  
For an eclipsing binary (second row in Figure~\ref{fig:all-views}), our representation includes the deep secondary eclipse at phase 0.5.  Hence, the 
Half-Phase view comprises the primary and a secondary eclipse.  In contrast, for a bona-fide
planet (first row in Figure~\ref{fig:all-views}) there is no secondary eclipse, and hence the Half-Phase view contains only the primary transit.

Our method differs from previously explored methods for constructing secondary views.
\cite{Yu_2019} construct a secondary view by searching the light curves for the mid-point of the most likely secondary eclipse (i.e., masking the transits
from the light curve and then using BLS to fit a box to find a transit) and then generate a binned/phase-folded view as input to their \texttt{Astronet-vetting} model. 
While this approach will likely identify secondary eclipses occurring anywhere within a period, a downside is that it may not generalize well.   For instance, planetary systems with multiple transiting
planets could be incorrectly classified as eclipsing binaries, adversely impacting model performance (i.e., low recall).   Hence, we chose to pursue a different approach.

\subsection{Stellar/Transit Parameters}
\label{subsec:stellar-transit}
 \cite{Ansdell_2018} show that providing additional 
domain features can improve model performance for deep learning based classifiers.  Therefore, 
similar to \cite{Osborn_2020}, we provide the following additional inputs to our model: (1) stellar parameters that we download from MAST catalogs 
namely stellar effective  temperature ($T_{\text{eff}}$), stellar log g ($\log g$), star radius ($R_*$), star mass, luminosity, stellar density,
and (2) transit parameters computed by TLS namely, transit depth, transit duration, $R_p/R_s$, mean odd/even transit depth.
Lastly, similar to \cite{Osborn_2020}, we normalize
each of these additional data columns by subtracting the median value and dividing by the standard deviation.

% next section: model
\section{\OurModel\ Pipeline}
\label{sec:models}
 We begin this section by providing an overview of our \emph{\OurModel\ } pipeline for identifying planetary candidates (\S~\ref{subsec:nigraha}).
 Next, we describe the
 neural network architecture
 of our machine learning model (\S~\ref{subsec:arch}).  
 We then
 describe our model training and prediction methodology (\S~\ref{subsec:training-method}).  We then evaluate model performance
 in (\S~\ref{subsec:eval}) and provide a brief comparison with prior
 approaches (\S~\ref{subsec:discuss}).
 
 %%%%%%%%%%%%%%%%%%%%%%%%%%%%%%%%%%%%%%%%%%%%%%%%%%%%%%%%
\subsection{\OurModel\ Pipeline Overview}
\label{subsec:nigraha}
\begin{figure}
 \begin{center}
\includegraphics[width=\columnwidth]{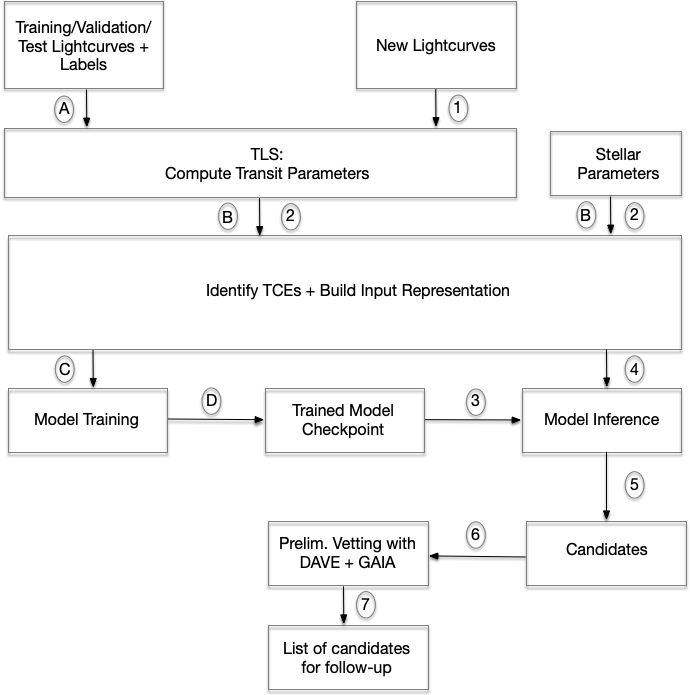}
\caption{\OurModel\ pipeline for model training, inference, and preliminary vetting of TCEs. }
\label{fig:e2e-pipeline}
\end{center} 
\end{figure}

Figure~\ref{fig:e2e-pipeline} depicts the training and inference (i.e., prediction) flows in our pipeline.
% The sequence of steps are:
% \begin{itemize}
% \item 
Steps (A)--(D) comprise the \emph{model training} path.
\begin{enumerate}[label=(\Alph*)]
\item The input light curves are processed using TLS which computes transit parameters. 
\item The transit parameters are  used to pre-process the TCEs to build the Global/Local/Half-Phase views.  Also,
the stellar parameters are combined to generate the input representation for the TCEs.
\item The input representations are then used for model training.
\item At the end of model training a checkpoint is generated which stores the model weights.
\end{enumerate}
%\item 
Steps (1)--(4) comprise the \emph{prediction} path.  Similar to the training path, Steps (1) and (2) generate the input representation for the TCEs.
Next, the model checkpoint is loaded (Step 3) and predictions are generated for each input (Step 4).  
The inference step generates a prediction (i.e., a score $\in (0, 1)$) for each TCE and this score ($PC_{score}$) is used to rank the candidates.
%{The inference step generates a prediction for each TCE that defines the likelihood of that TCE being caused by a planetary transit.}
We apply a threshold on $PC_{score}$ to choose candidates (Step 5).
%\item 
We perform preliminary vetting of the candidates (Step 6).  This is the final step of the pipeline and the output produced by
this step is a list of candidates for follow-up (Step 7).
%\end{itemize}

%%%%%%%%%%%%%%%%%%%%%%%%%%%%%%%%%%%%%%%%%%%
\subsection{Model Architecture}
 \label{subsec:arch}
\begin{figure}
 \begin{center}
\includegraphics[width=\columnwidth]{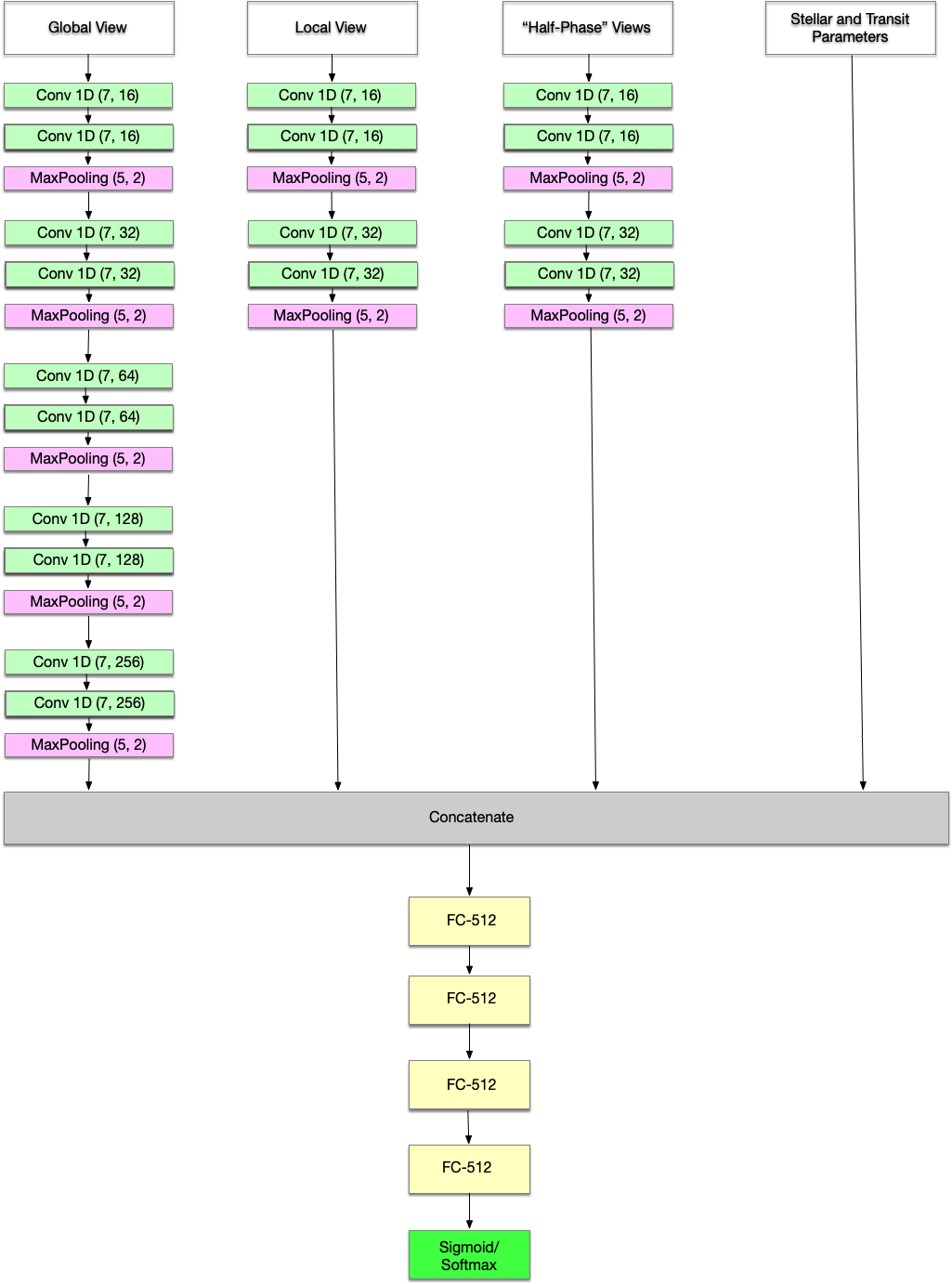}
\caption{Neural network architecture for our model.  Convolutional layers are
denoted as \texttt{Conv1D <kernel\_size, number of filter maps>}, pooling layers are denoted as 
\texttt{MaxPooling <pool\_size, stride\_size>}, and fully connected layers are denoted
as \texttt{FC-<number of units>}. The activation function in the output layer depends
on the type classifier---\texttt{sigmoid} for binary and \texttt{softmax} for multi-class classification. Currently we use only a binary model.
}
\label{fig:secondary-model}
\end{center} 
\end{figure}

 The \OurModel\ pipeline is built around a convolutional neural network based deep learning model.
 The architecture of our model is shown in Figure~\ref{fig:secondary-model}.  
 We use
 the \texttt{Astronet} architecture \citep{Shallue_2018} as the starting point for our model
 and the convolutional layers for global/local views is similar.  We then extend
 the model to include the half-phase view which has layers
 identical to the local view.  
 The Global, Local, and Half-Phase views are input as
 separate convolutional columns.  Similar to \texttt{Exonet} model \citep{Ansdell_2018}, the outputs of these convolutional layers are concatenated with the stellar and transit parameters and then
 passed to the fully connected layers.  We use the \texttt{relu} activation function for the convolutional layers.   Our model can be
 configured to perform either binary or multi-class classification.  For binary classification the output
 layer uses a  \texttt{sigmoid} activation function to generate a score in range (0, 1) that corresponds to the likelihood of the TCE being a planet candidate.  That is,
 values closer to 1 indicate a high likelihood of being a \texttt{PC} whereas values closer to 0 indicate that the TCE is likely to be a \texttt{FP}.  
 Extending our model to support multi-class classification is part of on-going work in this project and is beyond the scope of this paper.  We defer discussion to \S~\ref{sec:conclus}.

%%%%%%%%%%%%%%%%%%%%%%%%%%%%%%%%%%%%%%%%%%%%%%%%%%%%%%%%%
\subsection{Model Training and Prediction Methodology}
\label{subsec:training-method}
We implement our deep learning model using Keras \citep{chollet2015keras} on Tensorflow \citep{abadi2016tensorflow}.  We use
the following methodology to drive an evaluation.

\noindent \textbf{Model Training:} We train our models using the labeled datasets described in \S~\ref{subsec:input-rep}.  The labels corresponding to the positive and negative examples used in the training are defined in Table~\ref{tab:TESS-TCEs}.
For model training and evaluation, as is customary with training machine learning models, we
randomly split the labeled data into 3 sets: namely, 80\% is the \emph{training set} used for model training, 10\% is the
\emph{validation set} used for model validation, and 10\% is set aside as the \emph{test set} which is used solely for model evaluation.  Additionally, we augment the test set by injecting TCEs corresponding to known planets and eclipsing binaries (see \S~\ref{subsec:eval}).
We train our models using a batch size of 64 with the Adam optimizer \citep{kingma2014adam}.  We enabled early stopping in the training step and the models were trained for roughly 45-60 epochs.  We tuned a few of the  hyperparameters of the Adam optimizer
and found that the best models were with values for learning rate $\alpha = 1e-5$ and $\epsilon = 1e-8$. 

It is well known that model training with imbalanced distributions is hard as the model biases its learning towards features of the negative
examples \citep{Chawla2005DataMF}.  We took two steps to remedy this problem.  First, we initialize the output bias in the \texttt{sigmoid} output layer to 
the ratio of the positive to negative examples (i.e., $\log(pos/neg)$).  Second, we use balanced-batch sampling which is a well-known oversampling technique to train
the model.  With balanced-batch sampling, each mini-batch contains an equal number of positive and negative examples \citep{He_Garcia_2009}.

\noindent \textbf{Generating Predictions:} We use ``model averaging'' to generate predictions.  That is,
we train 10 independent models on the same training set using random parameter initialization and 
then compute the likelihood of the TCE being a planet as the average of the score predicted by each model.   
Model averaging is a well-known ensemble learning technique that is commonly used to minimize variance in a neural network model performance \citep{karpathy_2019}.

%%%%%%%%%%%%%%%%%%%%%%%%%
\subsection{Evaluation}
\label{subsec:eval}
Our test set consists of 632 TCEs of which
56 TCEs have one of \texttt{KP, PC, CP} labels  corresponding to the positive instances (i.e., only 8.8\% are positive).  To address the imbalance a little,
we augment the test set with an additional 175 labeled TCEs from Sector 6 and Sectors 21---26.   The labels
for these TCEs are from the TESS TOI catalog and were assigned based on manual vetting by the TESS team \citep{Guerrero_2020}.   This augmented data set consists
of (1) 98 TCEs that are positive instances which correspond to either previously known planets (i.e., \texttt{KP}) or planets that were confirmed through followup of
TESS candidates (i.e., \texttt{CP}), and (2) 77 TCEs that are negative instances which have false positive labels, namely, \texttt{FP, EB}.
With this augmentation, the combined test set now has \CombinedTestPositive\ positive instances.   

With imbalanced data sets, it is generally recommended that the desirable metrics for evaluating model performance are AUC and precision-recall curves \citep{davis_goadrich_2006}. We evaluate
these two metrics in the following subsections.  
Additionally, we examine how our model scores on the various TCE classes (\texttt{KP, CP, PC}) in the augmented test set. For this analysis, we also include the \texttt{PC}'s from the TOI catalog from Sectors 1--5 and 7--20. 

\subsubsection{AUC}
The AUC metric (area under the curve) corresponds to the area under the receiver-operator characteristic curve, which is equivalent to the probability that a randomly selected positive is ranked above a randomly selected negative.
The AUC score for our model on the combined test set is \AUCScore.

\subsubsection{Setting $PC_{score}$ threshold}

\begin{figure}
 \begin{center}
\includegraphics[width=\columnwidth]{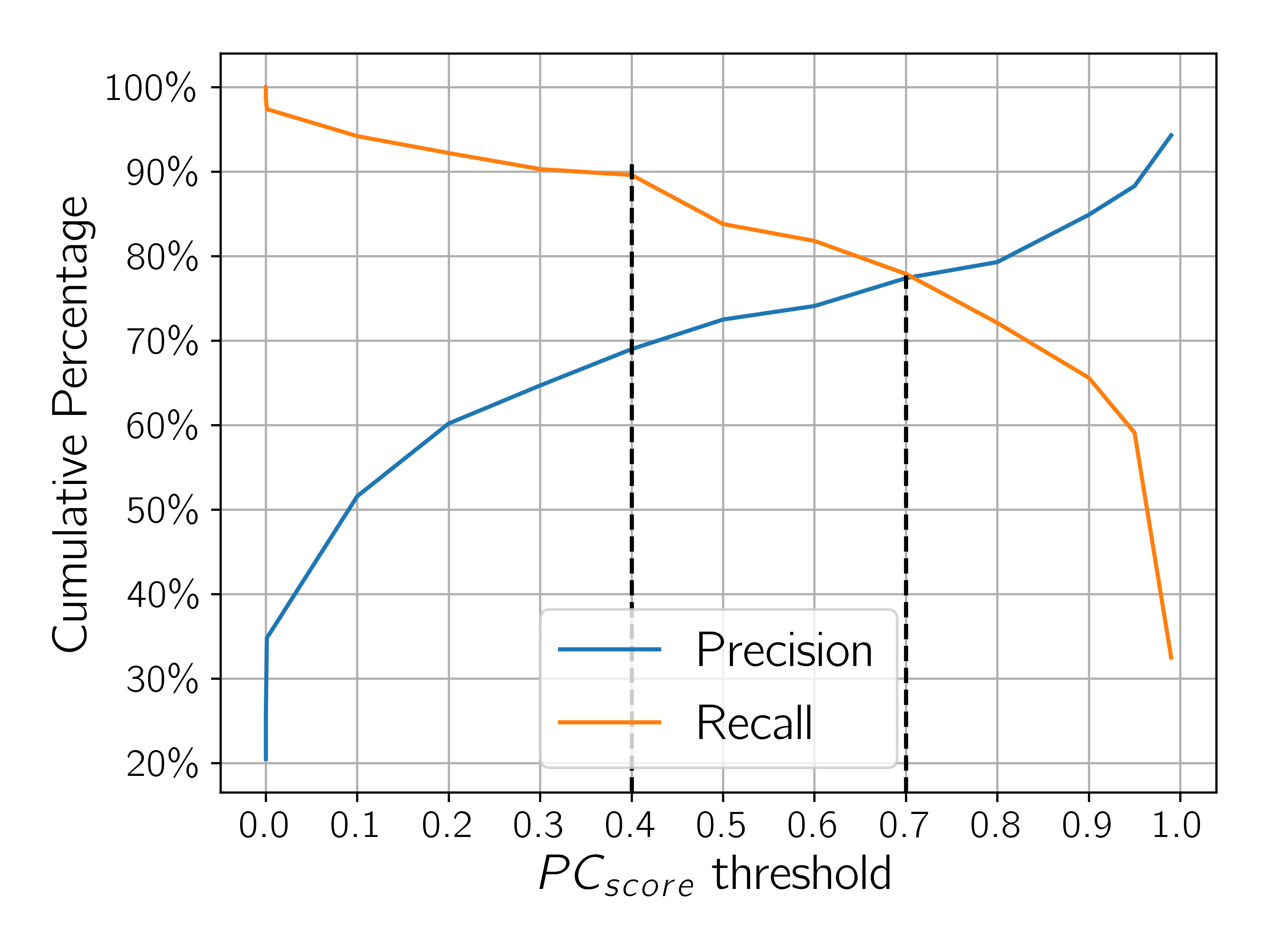}
\caption{Precision-recall curves as a function of $PC_{score}$.
Moving the threshold leads to a precision-recall trade-off.
} 
\label{fig:precision-recall-threshold}
\end{center} 
\end{figure}

Figure~\ref{fig:precision-recall-threshold} shows the precision-recall curves
as a function of threshold ($PC_{score}$) for the test set.  For the analysis in this section, 
we choose the threshold by picking a point closest to the intersection  of the two curves as it represents a trade-off between precision and recall. 
  As the figure shows, this point corresponds to a threshold of $PC_{score} > \PCprob$ and we use this for candidate selection.

\begin{figure}
 \begin{center}
\includegraphics[width=\columnwidth]{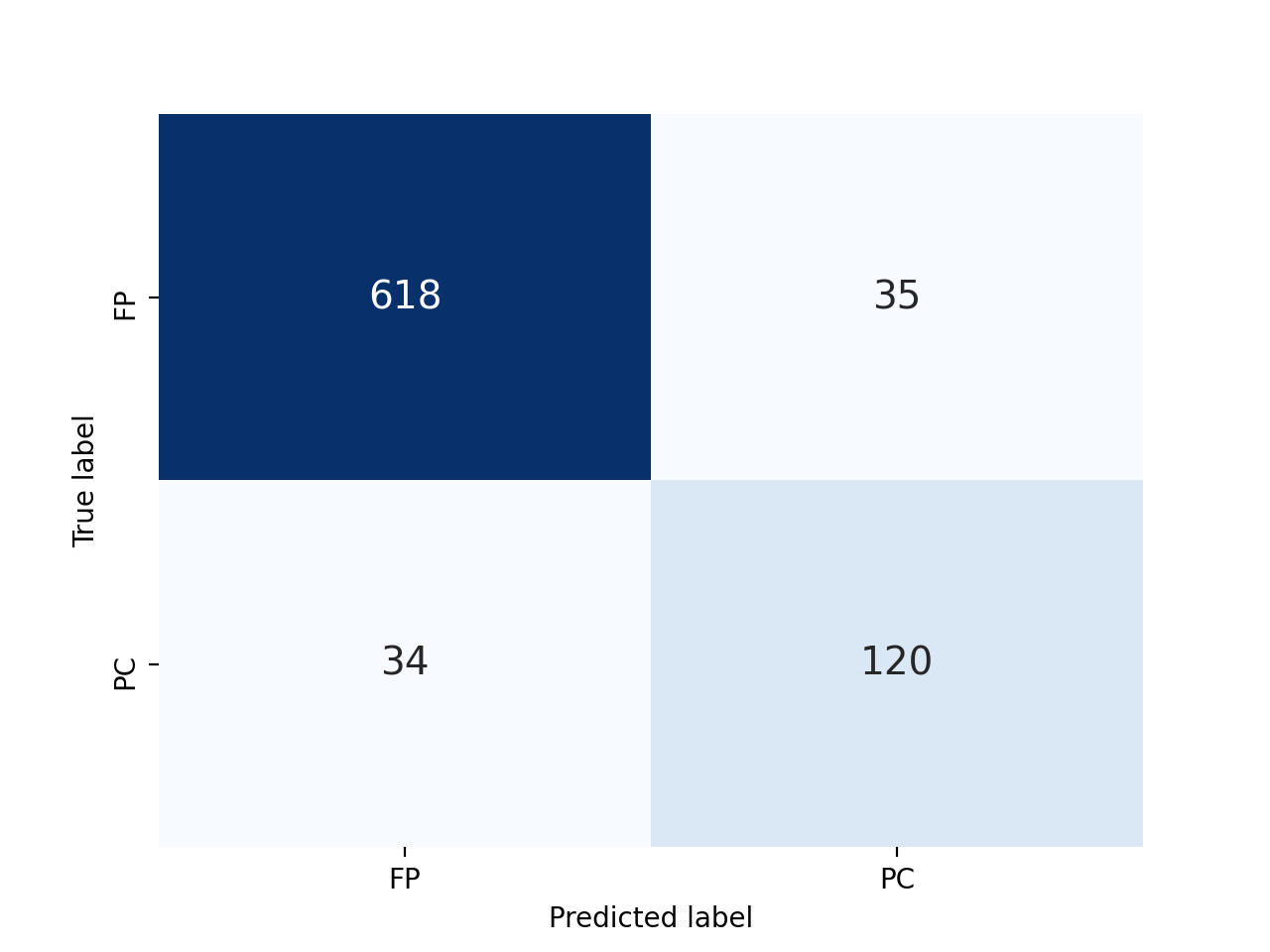}
\caption{Confusion matrix for our model where $PC_{score} > \PCprob$ is used to identify planet candidates. PC are planet candidates (i.e., positive instances), and FP are false positives (i.e., negative instances).} 
\label{fig:confusion-matrix}
\end{center} 
\end{figure}

We next construct a confusion matrix for the test set at this threshold (see Figure~\ref{fig:confusion-matrix}).  
In the figure, we use \texttt{PC} to denote TCEs corresponding to the positive instances and \texttt{FP} to denote the negative instances.
The model has an accuracy of \TestAccuracy\ (fraction of the instances that are correctly classified), a precision
of \TestPrecision\ (fraction of the positive instances that are indeed true positives), and a recall of \TestRecall\ (fraction of the true positives that are recovered).

\begin{figure}
\begin{center}
\includegraphics[width=\columnwidth]{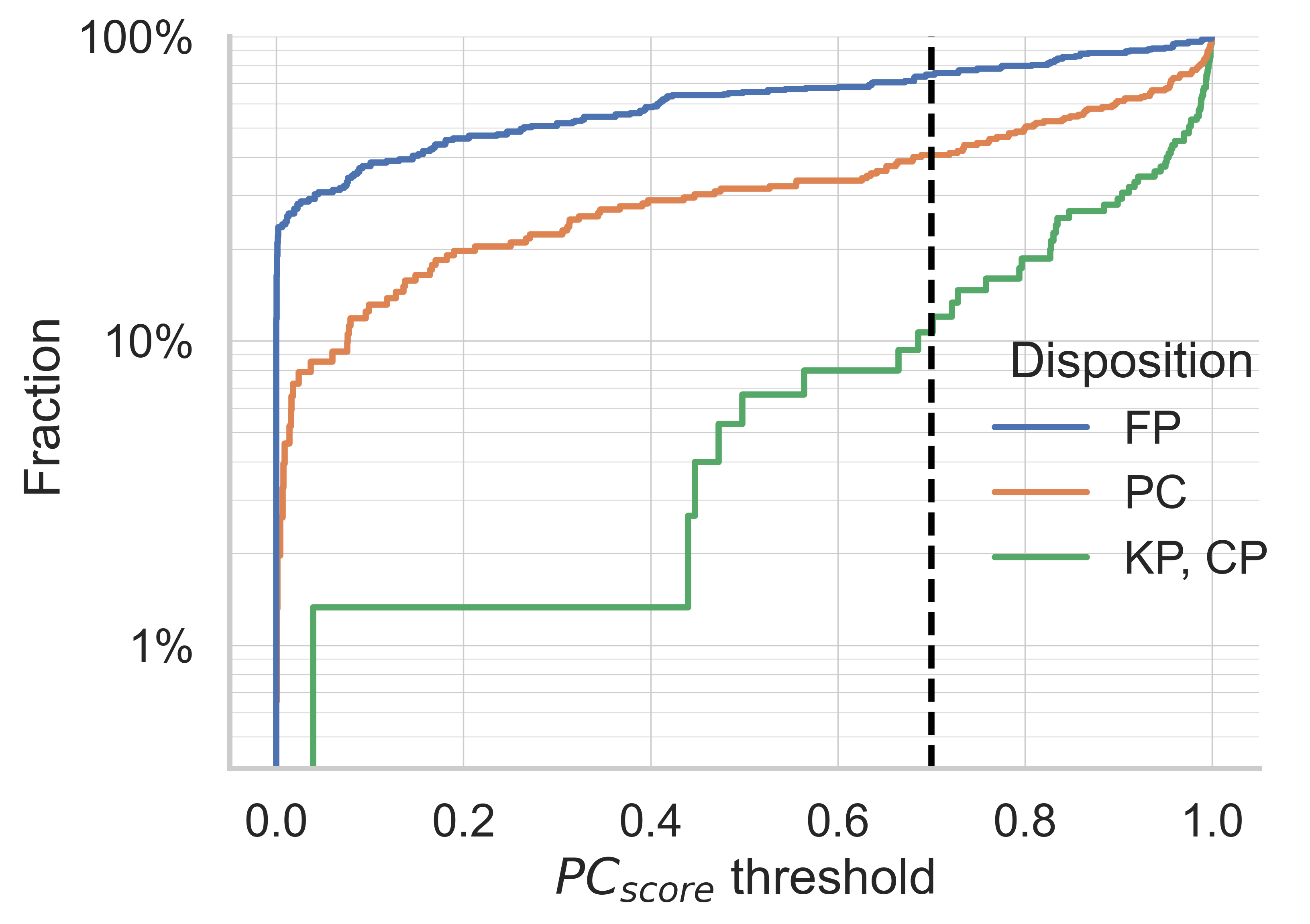}
\caption{A plot of the cumulative distribution function (CDF) that shows the proportion of TCEs scored by our model as a function of the $PC_{score}$ threshold.   In particular, as illustrated by the vertical dotted line, \OurModel\ assigns a high score ($PC_{score} > \PCprob$) to nearly 90\% of the bona-fide planets (\texttt{KP, CP}).} 
\label{fig:kp-cdf}
\end{center} 
\end{figure} 

%\begin{figure}
%\begin{center}
%\includegraphics[width=\columnwidth]{figures/toi-fp-snr-by-range-hist.png}
%\caption{A plot that shows the distribution of the $PC_{score}$ assigned by \OurModel\ to the false positive class of TCEs.  We bucket TCEs by their TLS computed SNR and classify TCEs with $TLS_{SNR} > 25$ as \emph{high}, $TLS_{SRN} < 16$ as \emph{low}, and the rest as \emph{med}.} 
%\label{fig:toi-fp-snr}
%\end{center} 
%\end{figure} 

%%%%%%%%%%%%%%%%%%%%%%%%%%%%%%%%%%%%%%%%%%%%%%%%%
\subsubsection{Model Scoring}
\label{subsec:threshold}
A question of interest is the $PC_{score}$ range our model assigns to TCEs corresponding to various classes.  
For this analysis, we also include previously classified \texttt{PC}'s from the TOI catalog.
Figure~\ref{fig:kp-cdf} shows the cumulative distribution function (CDF) of the fraction of TCEs
recovered above a given $PC_{score}$. Our model recovers \InjectionKP\ (\InjectionRecall) of the \texttt{KPs} and \texttt{CPs} with $PC_{score} > \PCprob$. Such a high recall at a high threshold is
desirable.  However, for the same threshold, over 30\% of the false positives also get included.  Such misclassified false positives can be eliminated by downstream vetting.

Given the threshold of \PCprob, we find that our model fails to recover 8 TCEs.  Of these,  6 are \texttt{KP}'s and 1 is a \texttt{CP} for which our model  predicts $PC_{score}$ in the range of 0.44 to 0.68.
We visually examined the light curves for the missing objects and did not find any systematic issues.  It is reasonable to expect that as the quality of training data improves these objects would be scored higher. 
A \texttt{CP} we fail to recover (\TIC{260128333}) is assigned a score of $PC_{score} = 0.039$.  For this object, the Group Comments column in the TESS TOI catalog observes that ``some transits may be spurious so period could be at longer interval''. 

\begin{figure}
\begin{center}
\includegraphics[width=\columnwidth]{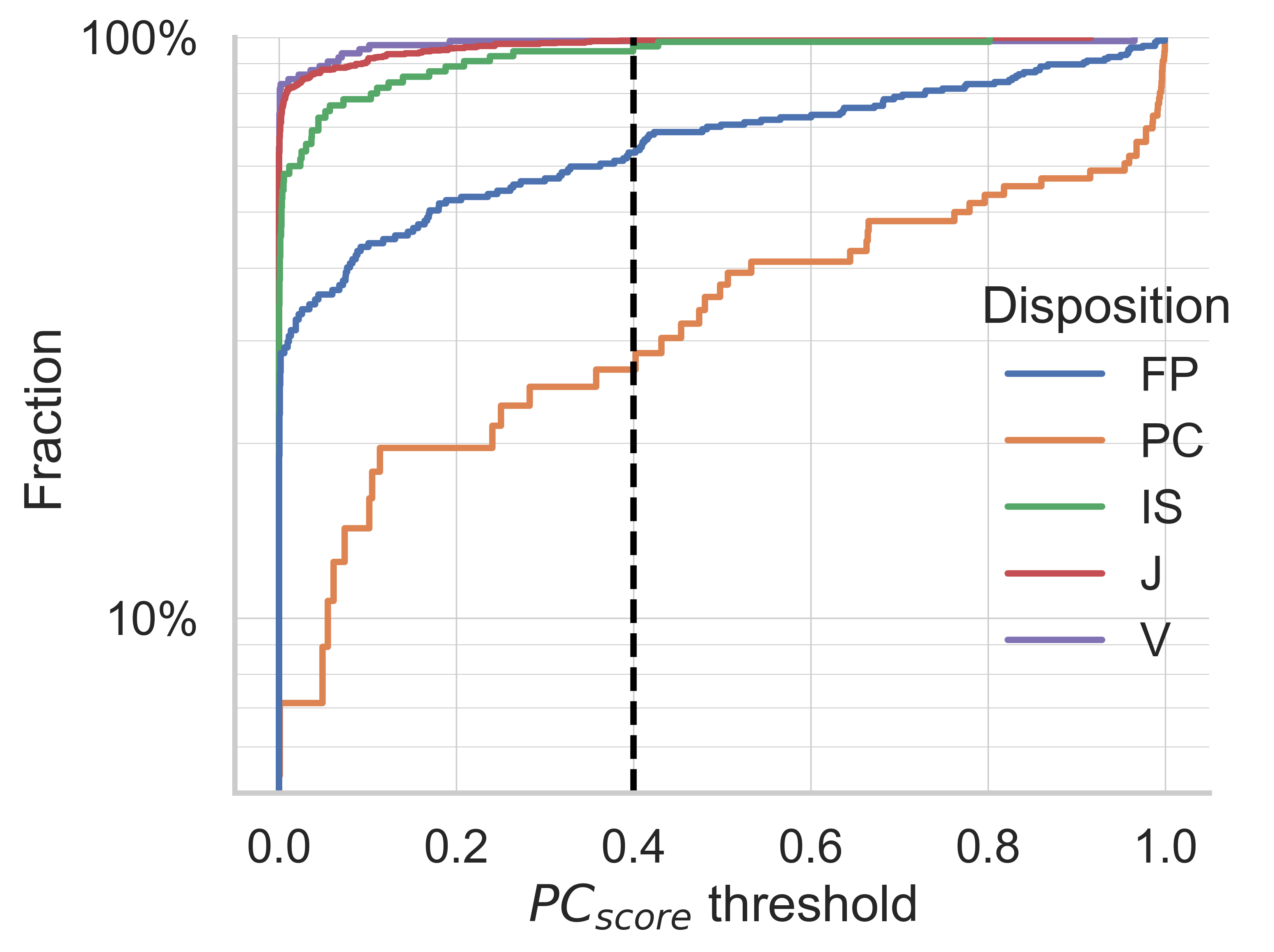}
\caption{A CDF plot that how the false positive TCEs are scored by our model as a function of the $PC_{score}$ threshold.  For comparison, we also include the CDF of the scores for the \texttt{PC} class.} 
%As illustrated by the horizontal dotted line, \OurModel\ assigns a low score ($PC_{score} < 0.4$) to nearly 97\% of the false positives.} 
\label{fig:fp-by-type-cdf}
\end{center} 
\end{figure}

Next, Figure~\ref{fig:fp-by-type-cdf} shows the CDF of the scores for the various classes of false positives in the test set.  On this figure, we also overlay the scores assigned to the \texttt{PC}'s from the test set.  Note that, the false positives are in general assigned much lower scores by \OurModel.  In particular, note that, for the \texttt{IS, V, J} classes, over 90\% of those TCEs are scored at 0.2 or less.

\begin{figure}
\begin{center}
\includegraphics[width=\columnwidth]{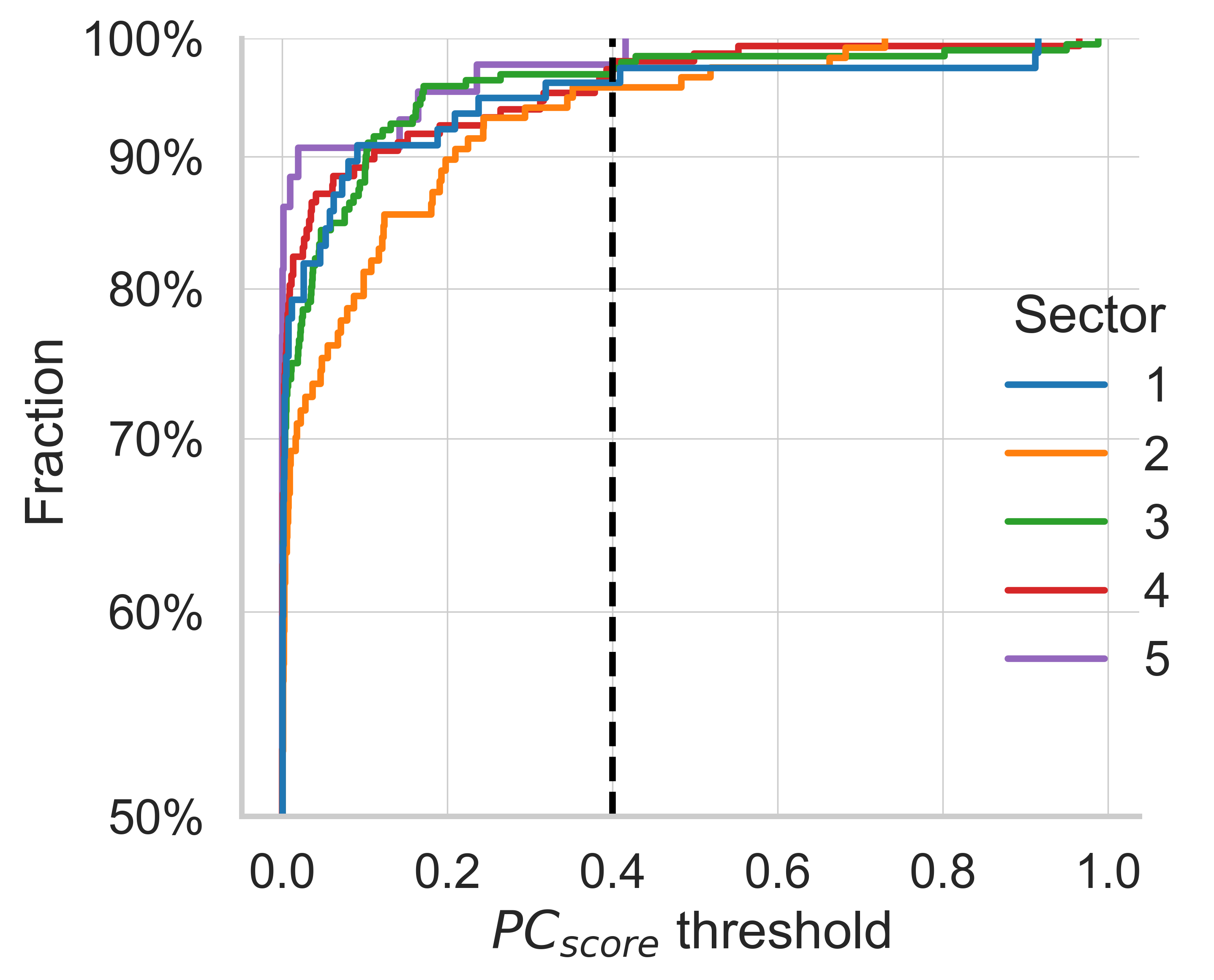}
\caption{A CDF plot that shows the score distribution as a function of the $PC_{score}$ threshold for the \texttt{IS, V, J} classes on a per-sector basis. As illustrated by the vertical dotted line, \OurModel\ assigns a low score ($PC_{score} < 0.4$) to nearly 95\% of the false positives.} 
\label{fig:fp-cdf}
\end{center} 
\end{figure}

Finally, as the noise properties and scattered light vary on per-sector basis, it is possible that \texttt{J, IS} may be sector dependent.  This is of interest as it could potentially affect our results: we train our model with data from Sectors 1--5 and then use that to predict on another sectors.  To understand this effect, Figure~\ref{fig:fp-cdf} shows the CDF of the scores for the
\texttt{IS, V, J} classes (from our test set) on a per-sector basis. As the graph shows, we find that across these sectors scores do not exhibit substantial variation.  That is, we find that nearly 95\% of the FP’s in each sector are scored 0.4 or lower. This is desirable as it suggests that \OurModel\ is able to consistently rule out vast majority of the false positives.

To summarize, picking the threshold is a precision-recall trade-off: it can be moved in either direction to increase or decrease recall or precision at the cost of the other depending on one's aim.  Thus, we could lower the threshold to \PCprobNew, and recover the missing \texttt{KP}'s, but at the expense of increasing false positives.  The false positives then will have to be excluded by the downstream vetting.

%%%%%%%%%%%%%%%%%%%%%%%%%%%%%%%%%%%%%%%%%%%%%%%%%%%%%%
\subsection{Comparision to Prior Models for TESS}
\label{subsec:discuss}

We find that \OurModel\ recovers 19 of the 22 TCEs with \texttt{KP} label from Sector 6 with  $PC_{score} > 0.8$.  In contrast \texttt{Astronet-Vetting} recovers 7 of these (see Table 2 of \citep{Yu_2019}). 
\cite{Osborn_2020} trained on simulated data and applied their model to TESS Sectors 1--4, whereas our training set included data from TESS Sectors 1--4.  Consequently, we can not do a direct comparison; but their recall is 61\% compared to \OurModel's \TestRecall. These non-exhaustive comparisons suggest that \OurModel\ generalizes better.

% next section: candidates
\section{Generating New Candidates}
\label{sec:candidates}
We use the \emph{inference} path of our pipeline (see Figure~\ref{fig:e2e-pipeline}) to identify candidates
from Sector 6 and Sectors 21---26.  
We begin by providing an overview how the various steps in the pipeline prune the candidates (Table~\ref{tab:cands}) and then describe our results.

\subsection{Overview}
Our TCEs
comprise the ones that have been found by the TESS Science Processing Operations Center (SPOC) pipeline \citep{2016SPIE.9913E..3EJ}, MIT Quick Look Pipeline (QLP) as well as previously
unseen TCEs.  The new TCEs we identify are due to the difference in the search procedure:  While we use TLS and it is better
at recovering shallow transits, the SPOC pipeline uses a wavelet-based adaptive matching filter for the search \citep{Jenkins_2010} 
and may have missed recovering those TCEs.   The effect of this difference is illustrated in Table~\ref{tab:cands} (Rows 2, 4, and 5).  
The inputs to both the pipelines is the same: Row 1 in the table
shows the number of unique light curves for the TESS targets across the various sectors.  From the light curves of these targets, Row 2 and Row 4, respectively, show the number of TCEs found
by SPOC and \OurModel.  In anticipation of the inputs needed by our ML model, we exclude objects that are missing at least one of the stellar parameters (described in \S\ref{subsec:stellar-transit}). Row 5 shows the TCEs that are common to both the pipelines. 

We use our trained model to generate a prediction (i.e., $PC_{score}$) for each of the \OurModel\ found TCEs.  
In this section, we apply a threshold $(PC_{score} > \PCprobNew)$ to identify a promising sample of candidates for further vetting. We use Figure~\ref{fig:kp-cdf} to pick this threshold for which there is a high recall (at least 90\%), 
and we use downstream vetting to eliminate false positives.
The candidate set we construct includes objects
that have been previously identified either by the TESS pipelines or by others in the community. Row 3 of Table~\ref{tab:cands} shows the number of labeled
objects across the various sectors that we found by
cross-referencing with the publicly available TESS TOI and TESS CTOI catalogs\footnote{\url{https://exofop.ipac.caltech.edu/tess/} accessed 10/12/20}.    Next, any candidates
that are left after we filter out
these known ones are \emph{new}.   Row 6 shows the count of the \emph{new} planetary candidates.
Lastly, we perform preliminary vetting (\S\ref{subsec:vetting}) using a series of steps including evaluation with \DAVE (Discovery and Validation of Exoplanets) \citep{Kostov_2019_DAVE} and cross-checking with GAIA
to rank our candidates.  The number of candidates subject to vetting is shown in Rows 7---10 of Table~\ref{tab:cands}.   The final confirmation of these candidates as bona-fide planets is only through specialized observations.  Hence, our aim with vetting is to minimize the number of candidates to the most promising ones.

\begin{table*}
    \centering
    \begin{tabular}{c|l|r|r|r|r|r|r|r|r|r}
    \hline
    \textbf{Row Number} &
    \textbf{Flow Step}  & \textbf{S6} & \textbf{S21} & 
    \textbf{S22} & \textbf{S23}
    & \textbf{S24} & \textbf{S25} & \textbf{S26} & \textbf{Total} \\ \hline
1 & TPS transit searches & 19995 & 19995 & 19999 & 19998 & 19992 & 19994 & 19992 & 82647\\
2 & SPOC TCE objects & 812 & 638 & 561 & 501 & 956 & 974 & 1012 & 3724\\ % 9074
3 & TESS (PC|FS) & 81|32 & 101|15 & 78|14 & 77|15 & 116|18 & 123|16 & 125|15 & 358|65 \\
%4 & \OurModel TCEs & 1275 & 1754 & 1197 &1083 & 1873 & 1748 & 1874 & 7945 \\ % 10804
4 & \OurModel\ TCEs & 710 & 771 & 608 & 571 & 928 & 1009 & 977 & 5574 \\
%5 & common TCE objects & 502 & 434 & 362 & 344 & 681 & 716 & 763 & 2526 \\ %3802
5 & common TCE objects & 388 & 340 & 275 & 278 & 604 & 603 & 663 & 3151\\
%6 & (New) Nigraha PC ($PC_{score} > 0.8$) & 25 & 18 & 9 & 13 & 29 & 40 & 29 & 119 \\ %163
%7 & Nigraha PCs with SPOC overlap  & 21 & 10 & 7 & 10 & 24 & 33 & 26 & 90 \\ %119
%8 & DAVE Vetting (PC/FP) & 6/15 & 2/8 & 2/2 & 1/9 & 1/18 & 4/14 & 6/3 & 21/69 \\
%9 & Gaia RUWE Vetting (PC/Bin) &  5/1 & 2/0 & 2/0 & 0/1 & 1/0 & 3/0 & 6/0 & 19/2 \\ 
%10 & Nigraha-only (PC/Bin) & 4/0 & 7/1 & 1/0 & 3/0 & 4/0 & 7/0 & 2/0 & 28/1\\
6 & \OurModel\ TCE ($PC_{score} > 0.4$) & 47 & 42 & 26 & 19 & 60 & 57 & 55 & 306/230  \\
7 & \OurModel\ TCE with SPOC overlap & 42 & 23 & 21 & 16 & 53 & 51 & 50 & 256/181 \\
8 & \DAVE Vetted (PC|FP) & 7|35 & 4|19 & 5|16 & 1|15 & 5|48 & 8|43 & 13|37 & 43|213 \\
9 & \OurModel-only TCEs  & 5 & 19 & 5 & 3 & 7 & 6 & 5 & 50 \\ % |49 unique \\
10 & \DAVE vetted \OurModel-only (PC|FP)  & 0|5 & 6|13 & 0|5 & 1|2 & 1|6 & 0|6 & 1|4 & 9|41 \\ \hline
    \end{tabular}
    \caption{The flow of numbers for sectors 6, 21---26, and deduplicated totals through the different stages of our pipeline.  
    (1) Starting from $\sim$20000 targets that all TESS sectors have, the numbers that remain after different steps are as follows: (2) SPOC/QLP found TCEs using BLS, (3) planet candidates  and false signals from followed-up SPOC TCEs (PC includes known planets and confirmed planets; FS includes false positives and false alarms), (4)  \OurModel TCE objects after excluding those with missing stellar parameters. These are more numerous than SPOC TCEs for some sectors likely due to TLS catching shallower transits, (5) the number of TCEs common to SPOC and \OurModel, 
    (6) New \OurModel\ PCs, (7) PCs with SPOC overlap (these have SPOC DV reports), (8) \DAVE vetting for the SPOC overlap PCs (FPs are false positives), (9) \OurModel-only PCs, (10), \DAVE vetting for \OurModel-only PC. The last column of rows 6 and 7 denotes sums from the other columns with and without duplicates.
    }
    \label{tab:cands}
\end{table*}
% https://tasoc.dk/info/docs.php has per sector details for TCEs etc.
% DAVE incorrectly declares a FP for systems with multiple planets.

%%%%%%%%%%%%%%%%%%%%%%%%%%%%%%%%%%%%
\subsection{Finding Candidates}

\begin{figure}
 \begin{center}
\includegraphics[width=\columnwidth]{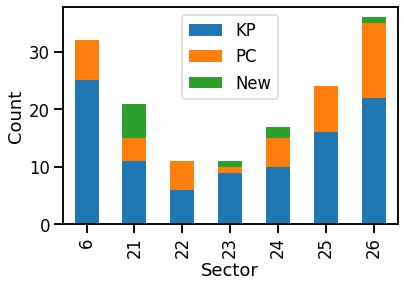}
\caption{We identify planet candidates based on the predictions generated by our model.  Since this includes
the previously known ones, the histogram provides a breakdown for each of the sectors we evaluated. \texttt{KP} are known planets, \texttt{PC} are planet candidates 
that are also common to SPOC (Table~\ref{tab:meta-vetted-pcs-1} and Table~\ref{tab:meta-vetted-pcs-2}), and \texttt{New} are vetted \OurModel-only planet candidates (Table~\ref{tab:meta-unvetted-pcs}) found by our ML pipeline.}
\label{fig:new-candidates}
\end{center}
\end{figure}

\begin{figure}
 \begin{center}
\includegraphics[width=\textwidth/2]{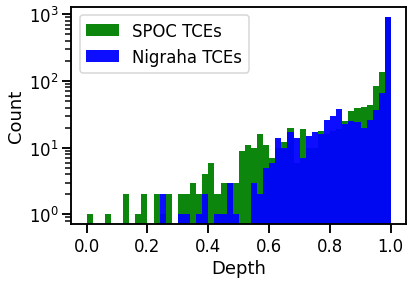}
\caption{SPOC and Nigraha TCEs. 
For the SPOC part we have used their transit depth in ppm - it varies ever so slightly with Nigraha depths.
Nigraha TCEs at shallower depths are far more numerous. Here and in the next two figures depth is described as the fraction of unobscured light.}
\label{fig:TCEs}
\end{center}
\end{figure}

\begin{figure}
 \begin{center}
\includegraphics[width=\textwidth/2]{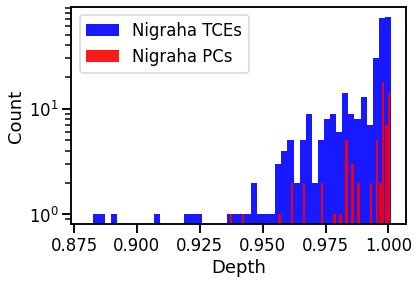}
\caption{Distribution of Nigraha TCEs (blue) which get filtered down to candidates (PC's, shown in red). For completeness depths from multiple sectors are included when same objects were found identified multiple times.}
\label{fig:NigrahaPCs}
\end{center}
\end{figure}

\begin{figure}
 \begin{center}
\includegraphics[width=\textwidth/2]{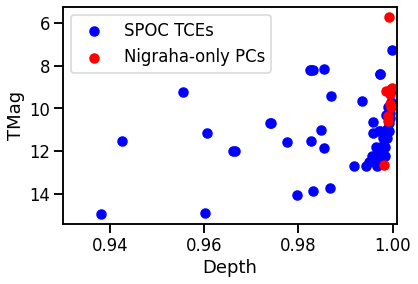}
\caption{Nigraha-only PCs (red) have shallower transits than SPOC TCEs (blue).}
\label{fig:NigrahaonlyPCs}
\end{center}
\end{figure}

Figure~\ref{fig:new-candidates} shows the number of TCEs in each of the sectors with a $PC_{score} > \PCprobNew$ as predicted by our model.  
The figure also provides a breakdown of the previously known planets (\texttt{KP+CP}) and candidates (\texttt{PC})
from the TESS TOI/CTOI catalogs.  After accounting for the known objects we are left with a candidate sample that is
roughly between 10 to 40 new candidates per sector (Row 6 of Table~\ref{tab:cands}).
Figure~\ref{fig:TCEs} shows the distribution of TCEs as a function of transit depth.   As expected, we find that \OurModel-only TCEs with
shallow transit depths to be numerous.  
Next, when apply our threshold for candidate selection ($PC_{score} > \PCprobNew$), the number of candidates reduces drastically.  This
is illustrated in Figure~\ref{fig:NigrahaPCs}.  This candidate sample includes TCEs found by both SPOC pipeline as well as \OurModel-only.  Figure~\ref{fig:NigrahaonlyPCs}
shows that \OurModel-only PC's tend to be fainter and are likely to have shallower transits compared to SPOC TCEs.
In what follows, we perform preliminary vetting of these candidates and then describe their characteristics.

%%%%%%%%%%%%%%%%%%%%%%%%%%%%%%%%%%%
\subsection{Preliminary Vetting Of New Candidates}
\label{subsec:vetting}

The next step in validating planet candidates as bona-fide planets is
to either perform
follow-up observations using telescope or use statistical methods.  Since telescope
time is scarce statistical validation is becoming an increasingly viable alternative. 
For instance, with Kepler/K2, \texttt{vespa}~\citep{Morton_2012, Morton_2016} has been extensively used to authoritatively validate hundreds of candidates as \emph{bona-fide} planets.  Such an
authoritative tool for TESS is not yet available.  
However, it is possible to do some preliminary vetting using \DAVE \citep[Discovery and Validation of Exoplanets;][]{Kostov_2019_DAVE}.  \DAVE which
was originally developed for K2 has been adapted and extensively used
for vetting TESS candidates \citep{Kostov_2019_L98-59,Gilbert_2020}.

\begin{figure*}
 \begin{center}
\includegraphics[width=\textwidth,]{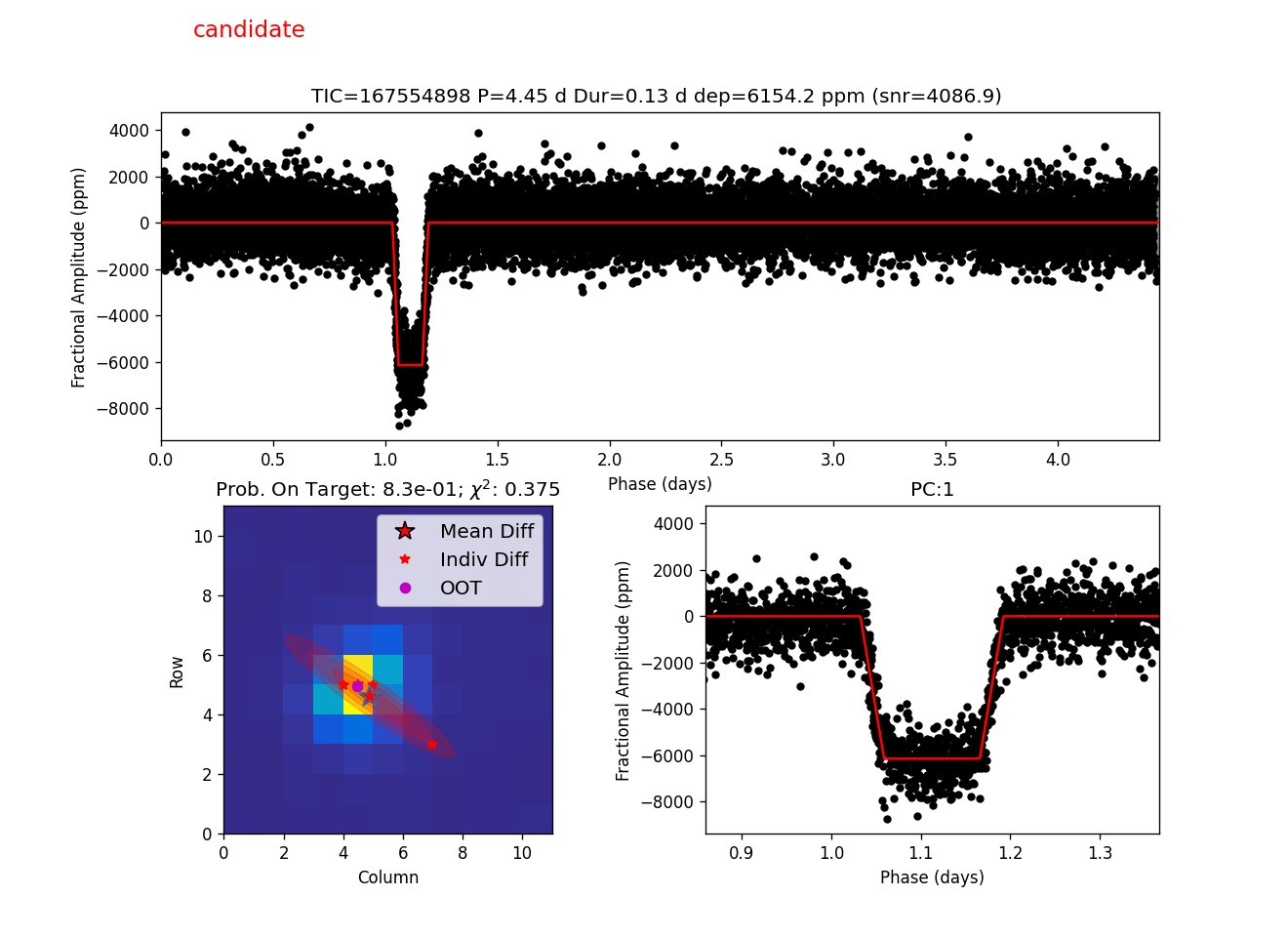}
\end{center}
\caption{Example summary report of vetting by DAVE. This is for our candidate \texttt{TIC 167554898}.}
\label{fig:dave-sample-output}
\end{figure*}

%\deleted{We perform preliminary vetting of these new candidates using \DAVE (Discovery and Validation of Exoplanets).  \DAVE  has been extensively used for vetting TESS candidates.}   
Briefly, \DAVE can be
configured to perform a set of tests to determine if the transit-like signals are due to common false positive scenarios: Using the data validation light curve, it checks for
odd-even differences between consecutive transits, secondary eclipses, stellar variability mimicking a transit; using the target pixel file, it checks for photocenter shifts during transit.
Both of these inputs to \DAVE are TESS products which are available for public download from MAST. 
Based on these tests, \DAVE provides a preliminary disposition: a TCE is either identified as a candidate or
it is flagged as a false positive with a diagnostic message.  For example, Figure~\ref{fig:dave-sample-output} shows the summary report for \texttt{TIC 167554898} which \DAVE\ identified as a candidate.

The data validation (DV) products (namely, light curves and reports) are available only for 
the TCEs identified by the SPOC pipeline.  
This is an issue for us since our method is capable of identifying TCEs not in the SPOC set (such as, those with shallow transits).
Hence, of the 243 candidates we identify, we were able to vet 211 TCEs across these sectors (line 7, Table~\ref{tab:cands}); 
of these,  \DAVE provided a positive disposition of \textsl{candidate} for 21 TCEs (line 8, Table~\ref{tab:cands}).  
The false positives consisted of: 
106 TCEs which failed the ModelShift test, 45 TCEs which failed the Odd-even depth test, and 63 TCEs which had centroid offset changes in the photometric image.
Additionally, there are instances where a TCE was classified by \DAVE as a candidate in one sector, but flagged as a false positive in another sector.  For such cases, we consider the TCE as a false positive. Conversely, there are instances where a TCE was classified by \OurModel\ as candidate in one sector but not in another.  For such cases, we do not consider the TCE as part of the candidate sample.

For these SPOC TCEs that are identified as a planet candidate by \DAVE\, we perform an additional vetting step.  For each target, we cross-check with the
associated SPOC generated data validation report as well as the TESS Exoclass catalog \citep{chrisburke2020_TEC}  to see if the diagnostic tests have flagged any issues such as, centroid shift, planet radius is too big, etc.
Table~\ref{tab:meta-vetted-pcs-1} lists the set of TCEs that pass both the vetting steps.   The diagnostic tests run as part of the SPOC data validation steps
are known to produce false negatives.  For instance, the data validation report for \TIC{103633434} flags a centroid out-of-transit test, but the TCE corresponds
to a \texttt{CP}.  Similarly, the TCEs from several other bona-fide planets in TOI catalog have been flagged as part of the data validation checks.  Hence, we report
those TCEs that did not pass the additional vetting separately and they are listed in Table~\ref{tab:meta-vetted-pcs-2}.  

We vet the remaining \OurModel-only candidates (line 9, Table~\ref{tab:cands}) with \DAVE by making a few modifications.  Since the DV products are not available for these targets, 
we input the \texttt{PDCSAP\_FLUX} light curves after detrending (as described in \S~\ref{sec:data}) along with our transit parameters to \DAVE.  Of the 51 \OurModel-only TCEs, we find that 3 pass all the tests and the remaining 48 are false positives.  These false positives consisted of: 37 TCEs which failed the ModelShift test, 10 TCEs which failed the Odd-even depth test, and 1 TCE which had centroid offset changes in the photometric image.  Table~\ref{tab:meta-unvetted-pcs} lists the set of \OurModel\ TCEs
that are identified as candidates by \DAVE.  

\begin{figure*}
 \begin{center}
\includegraphics[width=\textwidth]{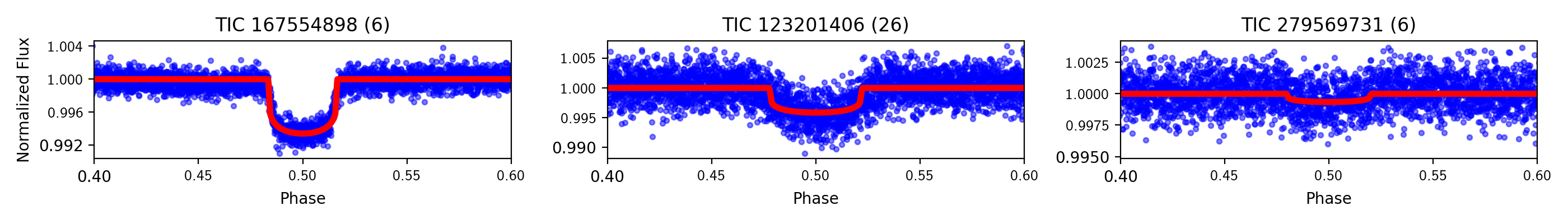}
\caption{The binned/phase folded light curves for the new planet candidates found by our model and subsequently vetted with \DAVE.
The SPOC TCEs in this panel pass all vetting tests and are not flagged in the DV report.}
\label{fig:vetted-all-good-candidates}
\end{center}
\end{figure*}

\begin{figure*}
 \begin{center}
\includegraphics[width=\textwidth]{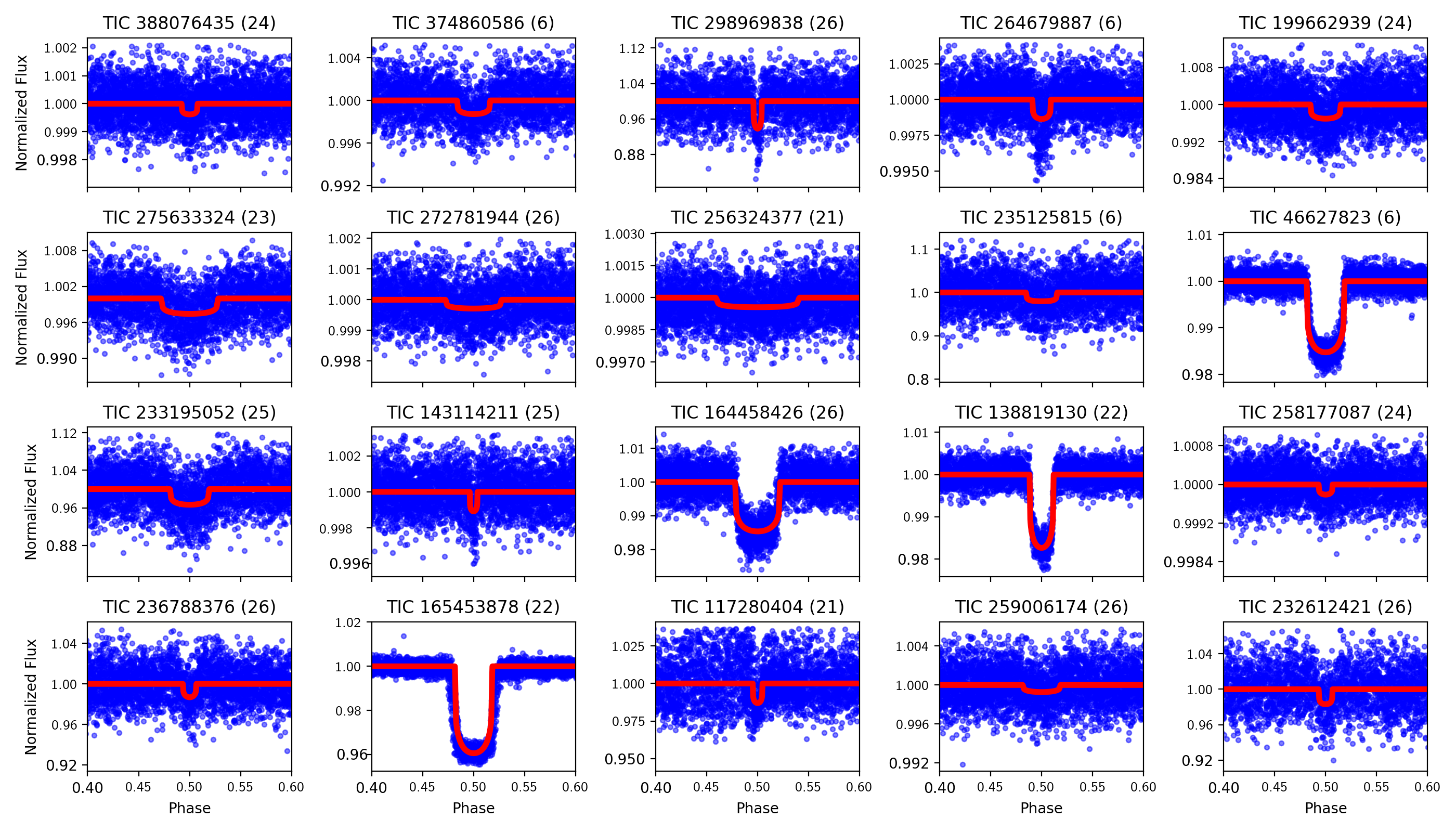}
\caption{The binned/phase folded light curves along with the transit model (shown in red) for the top-20 new planet candidates found by our model and subsequently vetted with \DAVE.
\emph{The SPOC TCEs in this panel have been flagged in the DV report.} }
\label{fig:vetted-candidates}
\end{center}
\end{figure*}

\begin{figure*}
 \begin{center}
\includegraphics[width=\textwidth]{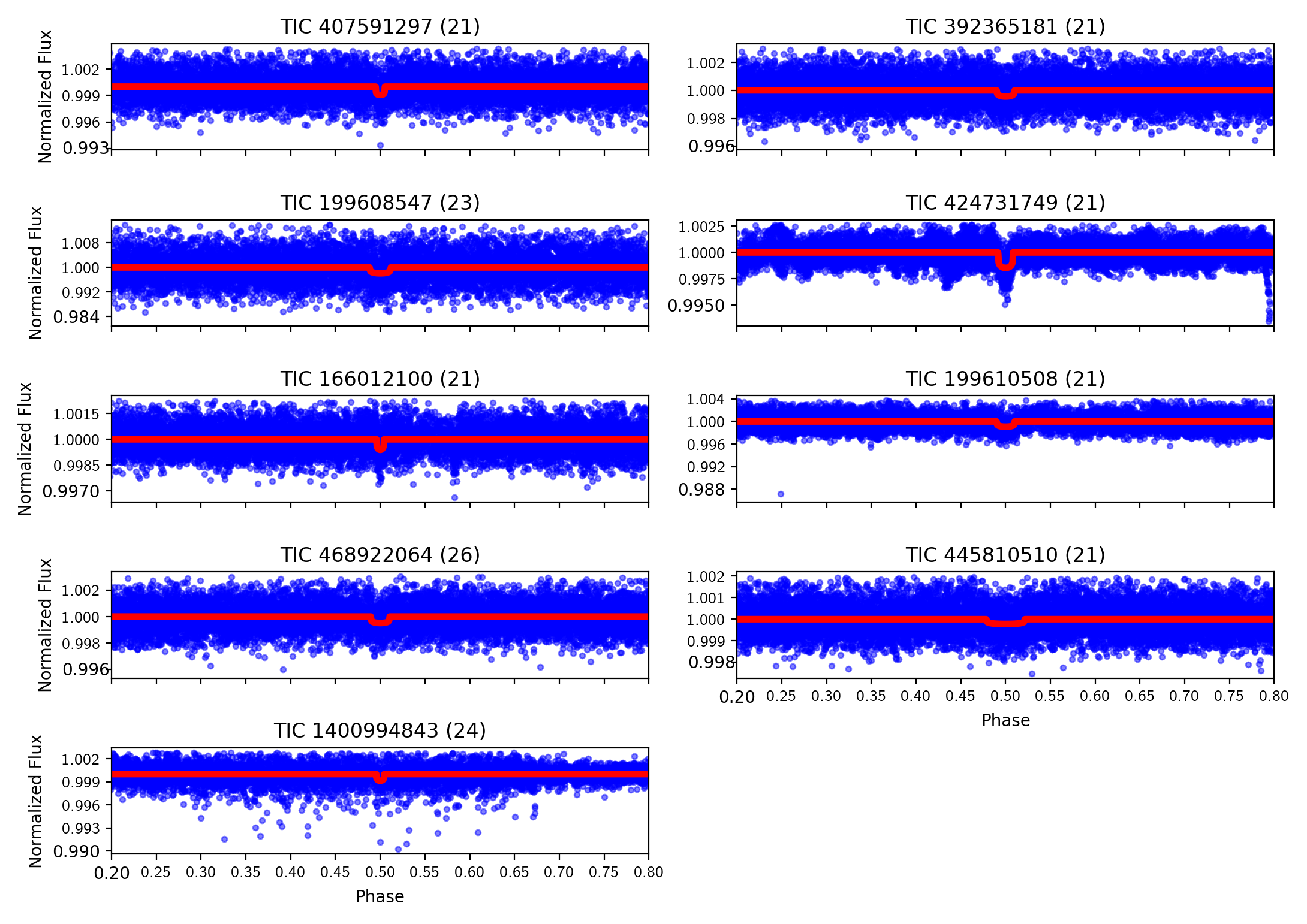}
\caption{
The binned/phase folded light curves along with the transit model (shown in red) for the new planet candidates found by our model.  The corresponding TCEs were not in the SPOC list.}
\label{fig:unvetted-candidates}
\end{center}
\end{figure*}

Figures~\ref{fig:vetted-all-good-candidates},
~\ref{fig:vetted-candidates}, and
~\ref{fig:unvetted-candidates}, respectively,
show the phase-folded light curves for all of the candidates from 
Tables~\ref{tab:meta-vetted-pcs-1},  ~\ref{tab:meta-vetted-pcs-2}, and ~\ref{tab:meta-unvetted-pcs}
which were identified by our pipeline.
We further checked in the Gaia catalog the Renormalised Unit Weight Error (RUWE) associated with each source. The RUWE is expected to be around 1.0 for single stars, and a value significantly greater, e.g. >1.4, 
possible binarity or an issue with the astrometric solution \citep{2018A&A...616A...2L}.

\begin{figure}
 \begin{center}
\includegraphics[width=\columnwidth]{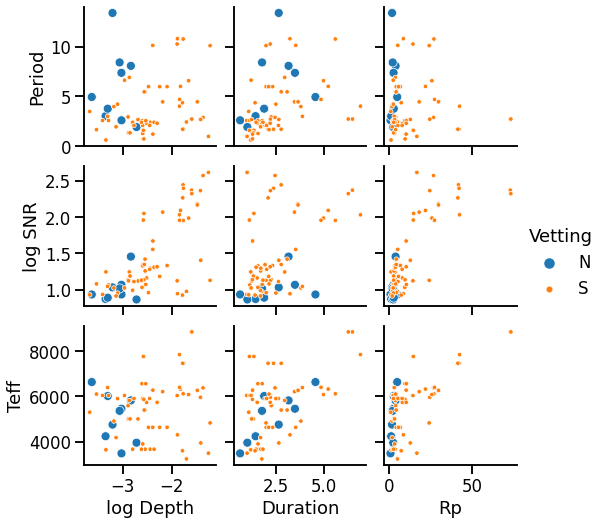}
\caption{Nigraha-only PCs (N) and those overlapping with SPOC (S).}
\label{fig:goodpps}
\end{center}
\end{figure}

Figure~\ref{fig:goodpps} shows some of the planetary characteristics for our vetted/unvetted candidate sample set.  In this figure, we compute a planet's radius
 as an approximation using the relation, $\delta_{\mbox{tra}} \approx k^2$, where $\delta_{\mbox{tra}}$ is the loss of light during transit and $k = \frac{R_p}{R_{\star}}$. 
 Note that, by using this approximation, we neither account for light contamination from nearby stars nor light from the planetary nightside which is typically negligible \citep{winn2014transits}.
 From Row 1 and Row 3 of this panel, we find that the final set of candidates we are left with are super-Earths ($R_p < 4 R_{\earth}$) with short period and duration, suggesting close proximity of their orbit to the host star.

%We intend to submit these new candidates to Exofop-TESS\footnote{\url{https://exofop.ipac.caltech.edu/tess/index.php}} as CTOIs for follow up.

\begin{small}
 \begin{table*}
 \begin{center}
 \begin{tabular}{rrrrrrrrrrrrr} \hline \hline
TIC/Sector & RA & DEC & TMag & Depth & Epoch & Duration & Period & $R_p $ & $R_{\star}$ & $PC_{score}$ & RUWE \\
  & & & & (ppm) &  & (hours) & (days)  & ($R_{\earth}$) & ($R_{\star}$) &   &  \\   \hline
  \color{red}{167554898/6} & 101.3158 & -61.0915 & 9.663 & 6579 & 1470.931 & 3.468 & 4.452 $\pm$  0.025 & \color{red}{18.320} & 2.068 &  0.811 &  1.29 \\
 123201406/26 & 280.9439 & 42.8392 & 11.165 & 4137 & 2010.889 & 1.262 & 1.203 $\pm$  0.004 & 10.478 & 1.492 &  0.609 &  1.34 \\
 279569731/6 & 105.9310 & -57.6608 & 10.442 & 654 & 1471.519 & 3.806 & 3.980 $\pm$  0.033 & 1.975 & 0.707 &  0.538 &  0.87 \\ \hline \hline
\end{tabular}
\caption{PCs from TCEs in Sctors 6 and 21---26 that overlap the SPOC TCEs. These are DAVE and Gaia vetted.  Various other metadata are provided to allow sub-selection for follow-up observing. $T_0 = \mbox{BJD} - 2457000$.  
Note that we do not account for light contamination from nearby stars when estimating planet radii.  Hence, our computed value is likely to be higher.  We take a conservative approach to flag all candidates whose planet radius exceeds $13$ by labeling them in red.}
 \label{tab:meta-vetted-pcs-1}
 \end{center}
\end{table*}
\end{small}

\begin{small}
 \begin{table*}
 \begin{center}
 \begin{tabular}{rrrrrrrrrrrrr} \hline \hline
TIC/Sector & RA & DEC & TMag & Depth & Epoch & Duration & Period & $R_p $ & $R_{\star}$ & $PC_{score}$ & RUWE \\
  & & & & (ppm) &  & (hours) & (days)  & ($R_{\earth}$) & ($R_{\star}$) &   &  \\   \hline
388076435/25 & 317.8838 & 68.4115 & 9.176 & 509 & 1984.819 & 1.117 & 2.576 $\pm$  0.007& 2.802 & 1.137 &  0.996 & 1.08 \\
 374860586/6 & 88.6250 & -62.6348 & 10.974 & 1279 & 1470.574 & 2.247 & 2.933 $\pm$  0.017& 4.181 & 1.070 &  0.993 & 0.84 \\
\color{red}{298969838/26} &288.5107 & 51.1625 & 14.965 & 61944 & 2019.446 & 1.970 & 10.115 $\pm$  0.056& \color{red}{24.324} &0.895 &  0.986 & 1.14 \\
264679887/6 & 82.1150 & 1.3003 & 10.305 & 1356 & 1468.840 & 2.956 & 6.914 $\pm$  0.038& 3.956 & 0.984 &  0.971 & 1.16 \\
\color{red}{275633324/23} &294.0727 & 57.2742 & 11.809 & 2563 & 1929.949 & 1.549 & 1.266 $\pm$  0.005& 3.527 & 0.638 &  0.895 & \color{red}{5.45} &\\
199662939/24 & 251.4267 & 54.5447 & 12.510 & 3075 & 1956.295 & 1.507 & 2.293 $\pm$  0.010& 8.197 & 1.354 &  0.895 & 1.09 \\
%167554898/6 & 101.3158 & -61.0915 & 9.663 & 6579 & 1470.931 & 3.468 & 4.452 $\pm$  0.025& 18.320 & 2.068 &  0.811 & \\
272781944/26 & 238.1813 & 68.7650 & 9.085 & 287 & 2011.568 & 2.131 & 1.652 $\pm$  0.010& 4.146 & 2.240 &  0.778 & 0.96 \\
\color{red}{256324377/21} &297.8002 & 62.2866 & 9.688 & 448 & 1870.822 & 1.167 & \color{red}{0.611 $\pm$  0.002} &1.350 & 0.584 &  0.730 & 0.93 \\
\color{red}{235125815/6} &102.1588 & 10.1983 & 14.053 & 20262 & 1469.040 & 1.749 & 2.423 $\pm$  0.013& 5.252 & 0.338 &  0.702 & \color{red}{4.63} &\\
\color{red}{46627823/6} &85.1816 & -17.5457 & 11.002 & 15258 & 1470.504 & 5.247 & 6.011 $\pm$  0.044& \color{red}{22.091} &1.638 &  0.700 & 1.12 \\
233195052/25 & 264.2525 & 67.0935 & 14.887 & 39835 & 1985.136 & 2.647 & 2.679 $\pm$  0.020& 12.957 & 0.594 &  0.668 & 0.99 \\
\color{red}{143114211/25} &257.0213 & 17.9605 & 9.924 & 1082 & 1988.147 & 1.185 & 6.636 $\pm$  0.018& 2.696 & 0.750 &  0.662 & \color{red}{4.07} &\\
\color{red}{164458426/26} &283.1663 & 45.4030 & 11.872 & 14640 & 2010.796 & 4.868 & 4.695 $\pm$  0.036& \color{red}{27.560} &2.086 &  0.647 & 0.97 \\
%123201406/26 & 280.9439 & 42.8392 & 11.165 & 4137 & 2010.889 & 1.262 & 1.203 $\pm$  0.004& 10.478 & 1.492 &  0.609 & \\
\color{red}{138819130/22} &175.4461 & 25.6577 & 11.508 & 17369 & 1909.557 & 5.623 & 10.781 $\pm$  0.113& \color{red}{27.187} &1.889 &  0.599 & 0.91 \\
258177087/24 & 239.6334 & 27.7400 & 7.290 & 208 & 1956.702 & 1.060 & 3.484 $\pm$  0.015& 1.212 & 0.769 &  0.564 & 1.03 \\
236788376/26 & 283.2101 & 56.0622 & 13.729 & 13243 & 2019.902 & 3.232 & 10.810 $\pm$  0.085& 8.810 & 0.701 &  0.555 & 1.00 \\
\color{red}{165453878/22} &200.8407 & 50.8268 & 11.165 & 39322 & 1902.021 & 2.202 & 2.706 $\pm$  0.005& \color{red}{24.694} &1.140 &  0.546 & 0.97 \\
%279569731/6 & 105.9310 & -57.6608 & 10.442 & 654 & 1471.519 & 3.806 & 3.980 $\pm$  0.033& 1.975 & 0.707 &  0.538 & \\
\color{red}{117280404/21} & 131.1200 & 28.4564 & 9.404 & 13059 & 1876.781 & 2.190 & 10.274 $\pm$  0.041& \color{red}{14.558} & 1.167 &  0.514 & 1.02 \\
259006174/26 & 290.3177 & 70.2361 & 11.057 & 725 & 2010.483 & 1.684 & 1.939 $\pm$  0.007& 2.064 & 0.702 &  0.457 & 0.79 \\
232612421/26 & 259.0272 & 63.1145 & 13.879 & 16792 & 2014.134 & 1.396 & 4.361 $\pm$  0.018& 7.499 & 0.530 &  0.447 & 1.05 \\
167696018/6 & 102.8952 & -62.7181 & 10.649 & 4122 & 1468.363 & 3.537 & 10.140 $\pm$  0.115& 5.426 & 0.774 &  0.441 & 0.88 \\
\color{red}{368260545/26} &265.4772 & 21.7358 & 11.587 & 22472 & 2015.478 & 4.997 & 6.578 $\pm$  0.050& \color{red}{25.965} &1.586 &  0.437 & 1.15 \\
468739039/25 & 20.3185 & 74.0719 & 10.208 & 485 & 1985.549 & 3.886 & 2.992 $\pm$  0.023& 3.290 & 1.367 &  0.423 & 1.00 \\
\color{red}{235726293/26} &276.4707 & 24.4778 & 11.508 & 57327 & 2011.717 & 0.975 & \color{red}{0.977 $\pm$  0.002} & \color{red}{16.929} & 0.647 &  0.417 & \color{red}{1.67} &\\
461634737/26 & 18.0993 & 88.3591 & 10.706 & 778 & 2011.987 & 2.491 & 4.220 $\pm$  0.031& 2.774 & 0.910 &  0.402 & 1.10 \\
\color{red}{233051157/26} &268.7985 & 61.1053 & 12.190 & 2693 & 2010.859 & 1.256 & \color{red}{0.720 $\pm$  0.003} &9.831 & 1.735 &  0.401 & 1.01 \\ \hline \hline 

\end{tabular}
\caption{PCs from TCEs in Sctors 6 and 21---26 that overlap the SPOC TCEs. Though these pass vetting by DAVE, they however fail some of the diagnostic tests that were run by the SPOC data validation pipeline.  Since those tests are known to produce false negatives, we list these PC's separately as many of these may still be viable for follow-up.  Also, we targets with a large radius ($> 13$), short period ($< 1 \mbox{day}$), and high RUWE score ($> 1.4$) are labeled in red.  Various other metadata are provided to allow sub-selection for follow-up observing. $T_0 = \mbox{BJD} - 2457000$.}
 \label{tab:meta-vetted-pcs-2}
 \end{center}
\end{table*}
\end{small}

\begin{small}
 \begin{table*}
 \begin{center}
 \begin{tabular}{rrrrrrrrrrrrr} \hline \hline
  TIC & RA & DEC & TMag & Depth & Epoch ($T_0$) & Duration & Period & $R_p $ & $R_{\star}$ & $PC_{score}$  & RUWE \\
 & & & & (ppm) &  & (hours) & (days)  & ($R_{\earth}$) & ($R_{\star}$) &   &  \\   \hline 
 %96202086/19 & 72.7559 & 31.4566 & 10.562 & 631 & 1816.209 & 0.405 & 0.630 $\pm$  0.001 & 0.987 & 0.360 &  0.965 & 1.16 \\
 407591297/21 & 152.1755 & 35.5470 & 10.574 & 936 & 1870.655 & 0.607 & 2.594 $\pm$  0.007& 1.046 & 0.313 &  0.859 & 1.16 \\
 392365181/21 & 155.9795 & 44.1149 & 9.954 & 441 & 1871.029 & 1.420 & 3.014 $\pm$  0.012 & 1.461 & 0.637 &  0.655 & 0.90 \\ % companion 5'' away
 199608547/23 & 248.7099 & 56.7093 & 12.656 & 1909 & 1931.255 & 0.988 & 1.917 $\pm$  0.007& 2.680 & 0.562 &  0.628 & 1.08 \\
424731749/21 & 255.1672 & 60.4345 & 9.183 & 1461 & 1871.163 & 3.159 & 8.078 $\pm$  0.037& 4.114 & 0.986 &  0.611 & 1.08 \\
\color{red}{166012100/21} &212.0799 & 64.0684 & 9.313 & 613 & 1870.525 & 2.640 & \color{red}{13.405 $\pm$  0.044} &1.995 & 0.738 &  0.600 & 0.95 \\
199610508/21 & 249.1159 & 58.5853 & 10.372 & 934 & 1876.602 & 3.495 & 7.375 $\pm$  0.079& 2.889 & 0.866 &  0.568 & 1.29 \\
\color{red}{468922064/26} & 96.6825 & 70.8227 & 9.779 & 488 & 2013.372 & 1.872 & 3.765 $\pm$  0.015& 3.031 & 1.255 &  0.553 & 1.11\\ 
\color{red}{445810510/21} & 170.4891 & 50.6882 & 9.067 & 229 & 1871.725 & 4.574 & 4.940 $\pm$  0.035& 5.092 & 3.077 &  0.457 & \color{red}{2.69}\\ 
1400994843/24 & 261.2504 & 67.3067 & 5.717 & 858 & 1958.469 & 1.761 & 8.421 $\pm$  0.049& 2.378 & 0.743 &  0.401 & *\\
\hline \hline

%1400994843/24 & 261.2504 & 67.3067 & 5.717 & 858 & 1958.469 & 1.761 & 8.421 $\pm$  0.049& 2.378 & 0.743 &  0.401 & \\

\end{tabular}
\caption{As in Table~\ref{tab:meta-vetted-pcs-1} but for \OurModel-only PCs.  These are vetted with \DAVE. $T_0 = \mbox{BJD} - 2457000$. \TIC{392365181} (Sector 21) has a companion 5'' away. It is over 3.7 mag fainter and with a RUWE score of 1.07, and unlikely to have contributed significantly to the variability of \TIC{392365181}. 
\TIC{468922064} (Sector 26) has a companion that 3.5 mag fainter with high RUWE a few arcsec away. The nearest Gaia object to \TIC{1400994843} is over 7 arcsec away and has a RUWE of 1.05.
}
 \label{tab:meta-unvetted-pcs}
 \end{center}
\end{table*}
\end{small}

\subsection{Discussion: BLS vs TLS}
\label{subsec:blsvtls}

Some of the new candidates we identified above are from lightcurves that were not identified as containing TCEs by the SPOC pipeline.  We believe the underlying reason that the TCEs were missed is primarily due to the low SNR/SDE in the transit search with BLS.  
Note that the widely used thresholds for BLS are $SNR_{\mbox{BLS}} > 7.1, SDE_{\mbox{BLS}} > 9$ \citep{Vanderburg_2016, Crossfield_2016}.  
Table~\ref{tab:BLSvsTLS} shows the results of our analysis.  We find that 6 of the 9 \OurModel-only TCEs were not recovered by BLS.

Interestingly, we find that \TIC{388076435} is recovered as a candidate by \OurModel in Sector 24, whereas SPOC does not.  
 In this sector, our TLS search returned a depth of 383.76 ppm at a SNR of 8.6 and SDE of 11.8.   However, we find that the same TIC is also identified as a TCE by SPOC in Sector 25 
 (see Table~\ref{tab:meta-vetted-pcs-2})  with a depth of $562.4 \pm 66.5$ ppm (as per the DV report), while our TLS search returned 509 ppm.

\begin{table}
    \centering
    \begin{tabular}{l|l|r|r|c|c}
    \hline
    \hline
      \textbf{TIC} & \textbf{Sector}  & \multicolumn{2}{c|} {BLS} & \multicolumn{2}{c|} {TLS}\\
    & & \textbf{SNR} & \textbf{SDE}  &\textbf{SNR} & \textbf{SDE} \\   \hline
407591297 & 21 & 7.487 & 13.429 & 8.610 & 17.853 \\
392365181 & 21 & 9.223 & 9.672 & 7.458 & 13.388 \\
199608547 & 23 & 4.525 & 9.387 & 7.344 & 9.923 \\
424731749 & 21 & 26.640 & 9.355 & 28.623 & 10.167 \\
166012100 & 21 & 7.036 & 4.723 & 10.774 & 9.590 \\
199610508 & 21 & 13.252 & 6.296 & 11.700 & 9.149 \\
468922064 & 26 & 6.449 & 10.912 & 7.775 & 10.406 \\
445810510 & 21 & 7.552 & 7.810 & 8.606 & 11.531 \\ 
1400994843 & 24 & 35.305 & 8.983 & 10.424 & 11.368 \\ \hline \hline
    \end{tabular}
    
    \caption{Comparison of SNR/SDE for the \OurModel-only TCEs between BLS and TLS. 
    }
    \label{tab:BLSvsTLS}
\end{table}

% next section: summary
\section{Concluding Remarks}
\label{sec:conclus}
In large-scale sky surveys for detecting exoplanets, it is well-known
that the number of bona-fide planetary candidates is dwarfed by the number of false positives in the observed data.
For instance, with TESS, the expected yield is about 80 candidates per TESS sector\footnote{\url{https://tess.mit.edu/toi-releases/toi-release-faqs/}}
which comprises 20000 targets.   Using machine learning techniques to build models that enable rapid classification has proven to be
a promising approach \citep{Shallue_2018, Ansdell_2018,2019MNRAS.483.5534S, Yu_2019, Osborn_2020}.  The TESS team operates multiple pipelines (namely, SPOC and QLP) to identify promising candidates, several of which have
been validated as bona-fide planets.  

As a complementary effort, we set out to build a machine-learning based pipeline, \emph{\OurModel} using novel approaches from ground-up.  First, the method used to search for transit signals in a lightcurves (TLS)
differs from prior approaches (such as, MIT QLP for TESS) which use BLS and its variants.  Compared to BLS, TLS has better sensitivity towards recovering shallow transits
caused by near-Earth size planets.  Consequently, this expands the search space for identifying candidates.   Second, to enable the model to
better distinguish planetary candidates from false positives caused by eclipsing binaries,  
we use a novel input representation of the light curve.   We trained our model using data from TESS sectors and evaluated it.  
\OurModel\ has an AUC score of \AUCScore.
In a test comprising light curves from known and TESS confirmed planets, we find that our model recovers \InjectionKP\ known/TESS confirmed planets.  We then used our trained model to generate predictions for data from various TESS sectors.
Besides recovering known planets and planet candidates, our model identified new candidates.  We performed preliminary vetting of all candidates using a sequence of steps
including \DAVE to rule out false positive scenarios and checked for binarity using the RUWE parameter from GAIA. 
Although neither of these methods are foolproof, by minimizing the number of candidates we make them better candidates for followup.
This is true for the shallow candidates we discovered as well as the existing planet candidates we independently affirm. 
The value of finding vetted shallower transits is demonstrated by \TIC{388076435}. \OurModel\ finds it in Sector 24, and only using data in Sector 25 does it become a SPOC candidate. 
Our Tables \ref{tab:meta-vetted-pcs-1}, \ref{tab:meta-vetted-pcs-2}, \ref{tab:meta-unvetted-pcs} summarize the best cases in sectors 6, 21-26 including a few with reservations.

Some of these may yet turn out to be false positives (possibly diluted eclipsing binaries), but a subset of those will likely be binaries with very faint companions, perhaps even white dwarfs, and thus interesting in their own right.
In the near future we will be extending our model to support multi-class classification with eclipsing binaries being an explicit class, and variables another.

Our procedure for finding TCEs has been intentionally conservative, and there are a few ways to strengthen our approach to increase the candidate yield.    
For instance, we only search for a single TCE per light curve;  we process light curves on a per-sector basis; we discard
candidates for which the stellar/transit parameters are not available.  We intend to address these issues as part of on-going work in this project.  Furthermore, this should also help with
identifying both stars with multi-planet systems as well as planets with longer periods (such as, those with periods greater than 27 days and discovered by combining data from multiple sectors).  
We are also exploring ways to improve model performance based on ideas such as,
model training using data from Sector 6 and later as that data has improved camera pointing systematics.
There are also alternate input data representation and machine learning model architecture optimizations to be tried. 

We have released the code and models developed for this paper as open source\footnote{\url{https://github.com/ExoplanetML/Nigraha}}.  We intend to submit the candidates found by \OurModel\ pipeline to the TESS Exoplanet Followup Program (ExoFOP-TESS).
Independently we are proposing high-resolution, time-resolved photometry for the most promising of our candidates using WIRC with the Engineered Diffuser \citep{Stefansson2017,Vissapragada2020}
and dual-band CHIMERA \citep{Harding2016}
at the Palomar 5m telescope to rule out false positives and confirm the targets as planets.
We hope other teams will also observe these candidates.

\section*{Acknowledgements}
We thank the anonymous referee whose detailed feedback greatly improved this paper.
We thank David Ciardi, Joe Ninan Philip, and Shreyas Vissapragada for insightful comments and feedback on various aspects of this work.
We also thank Tim Morton for discussions and feedback on early drafts of this paper. 

Funding for the TESS mission is provided by NASA’s Science Mission directorate.
We acknowledge the use of public TESS Alert data from pipelines at the TESS Science Office and at the TESS Science Processing Operations Center.
This research has made use of the Exoplanet Follow-up Observation Program website, which is operated by the California Institute of Technology, under contract 
with the National Aeronautics and Space Administration under the Exoplanet Exploration Program.
This paper includes data collected by the TESS mission, which are publicly available from the Mikulski Archive for Space Telescopes (MAST).

This research made use of \textsc{Lightkurve}, a Python package for Kepler and TESS data analysis \citep{2018ascl.soft12013L}.

This work has made use of data from the European Space Agency (ESA)
mission {\it Gaia} (\url{https://www.cosmos.esa.int/gaia}), processed by
the {\it Gaia} Data Processing and Analysis Consortium (DPAC,
\url{https://www.cosmos.esa.int/web/gaia/dpac/consortium}). Funding
for the DPAC has been provided by national institutions, in particular
the institutions participating in the {\it Gaia} Multilateral Agreement.

AM acknowledges support from the NSF (1640818).
%%%%%%%%%%%%%%%%%%%%%%%%%%%%%%%%%%%%%%%%%%%%%%%%%%
\section*{Data Availability}

The data underlying this article is available for public download on the GitHub repository at \url{https://github.com/ExoplanetML/Nigraha}.  
Additionally, on the same GitHub repository, we have released the code and models developed as part of this work.

%%%%%%%%%%%%%%%%%%%%
%\section*{Additional Figures}

% \begin{figure}
%  \begin{center}
% \includegraphics[width=\textwidth/2]{figures/spoc_TCEs.png}
% \caption{SPOC TCEs}
% \label{fig:SPOC_TCEs}
% \end{center}
% \end{figure}



\bibliographystyle{mnras}
\bibliography{main} % if your bibtex file is called example.bib

%\begin{thebibliography}{99}
%\end{thebibliography}

%%%%%%%%%%%%%%%%%%%%%%%%%%%%%%%%%%%%%%%%%%%%%%%%%%

%%%%%%%%%%%%%%%%% APPENDICES %%%%%%%%%%%%%%%%%%%%%

%%%%%%%%%%%%%%%%%%%%%%%%%%%%%%%%%%%%%%%%%%%%%%%%%%

% Don't change these lines
\bsp	% typesetting comment
\label{lastpage}
\end{document}